\documentclass[%
 reprint,
 amsmath,amssymb,
 prd,
]{revtex4-2}
\pdfoutput=1
\usepackage{graphicx}
\usepackage[utf8]{inputenc}
\usepackage{lineno}

\usepackage{color}
\usepackage{xcolor}

\begin{document}

\title{Complexity of Magnetization and Magnetic Simplification}

\author{Daniel \'Avila}
 \email{daniel.avila@correo.nucleares.unam.mx}
\affiliation{Departamento de Física de Altas Energias, Instituto de Ciencias Nucleares,
Universidad Nacional Autónoma de México Apartado Postal 70-543, CDMX 04510, México} 
\author{César Díaz}
 \email{cadh@ciencias.unam.mx}
 \affiliation{Departamento de F\'isica, Facultad de Ciencias, Universidad Nacional Aut\'onoma de M\'exico, A.P. 70-542, Ciudad de M\'exico 04510, M\'exico}
\author{Leonardo Pati\~no}
 \email{leopj@ciencias.unam.mx}
\affiliation{Departamento de F\'isica, Facultad de Ciencias, Universidad Nacional Aut\'onoma de M\'exico, A.P. 70-542, Ciudad de M\'exico 04510, M\'exico}

\begin{abstract}
We use the Complexity=Volume (CV) prescription to study the effect of a magnetic field on the computational complexity for states in the gauge theories dual to two different gravitational models. In one of these theories the complexity increases with the intensity of the magnetic field, while in the other a more interesting behavior is discovered, resulting in a phenomenon that we term magnetic simplification. The relevant difference between the two theories is that the content of the second includes a scalar operator with a non-vanishing vacuum expectation value. This leads us to conclude that the direct impact of the magnetic field is to increase the complexity of a state, but it can indirectly lower it by diminishing the complexity associated to additional degrees of freedom when these do not vanish across the space. We additionally compare the results obtained working in the full ten dimensional backgrounds and in their effective five dimensional truncations, exhibiting that the question is still current about which surface, whether the uplift of the 5D extremal hypersurface or the extremal surface in 10D, should be used in the CV prescription.
\end{abstract}


\maketitle
\section{Introduction}
\label{Intro}

Recent studies on the emergence of spacetime, in the context of the AdS/CFT correspondence \cite{Maldacena:1997re}, have relied on the geometrization of quantum information quantities. Examples of this include the entanglement entropy, whose holographic dual is the area of an extremal surface in the bulk \cite{Ryu:2006bv,Hubeny:2007xt,Lewkowycz:2013nqa,Dong:2016hjy}, the entanglement of purification, which is dual to the area of the minimal crosssection of the entanglement wedge \cite{Takayanagi:2017knl}, and the computational complexity \cite{Susskind:2014rva,Stanford:2014jda,Brown:2015bva,Brown:2015lvg,Carmi:2016wjl}(or quantum circuit complexity), which is the main focus of this manuscript. Roughly speaking, the complexity $\mathcal{C}$ of a given state $|\psi\rangle$ is the minimum number of quantum gates required to produce said state from a particular reference state $|R\rangle$.

There are two main holographic candidates to be dual to the computational complexity of the gauge theory state (although recently it has been argued that there are many other possible candidates \cite{Belin:2021bga,Belin:2022xmt}). The first one is the ``Complexity=Action'' (CA) conjecture \cite{Brown:2015bva,Brown:2015lvg,Carmi:2016wjl}, which relates the complexity to the on-shell action of the gravitational theory evaluated in a bulk region known as the WheelerDeWitt patch. The second one is the ``Complexity=Volume'' (CV) conjecture \cite{Susskind:2014rva,Stanford:2014jda}, in which the complexity of the state is related to the volume of a certain extremal region in the bulk. More precisely, if we are interested in the computational complexity $\mathcal{C}$ of a given gauge theory state $|\psi(\tau)\rangle$ at a time $\tau$ we can obtain it from the expression
\begin{equation}
\mathcal{C}(|\psi(\tau)\rangle)=\text{max}_{\Sigma}\frac{\text{Vol}(\Sigma)}{G_{N}L},
\label{CVConjecture}
\end{equation}
where $\Sigma$ is a codimension-1 hypersurface in the bulk that intersects the boundary in the timeslice $\tau$, $G_{N}$ is Newton constant and $L$ is an arbitrary length scale which we will take to be the AdS radius. 

Much progress has been made to better understand both recipes for holographic complexity, such as studying the time evolution of the complexity \cite{Carmi:2017jqz,Swingle:2017zcd,Mahapatra:2018gig,Alishahiha:2018tep,HosseiniMansoori:2018gdu,Auzzi:2022bfd} and its relation to the so-called Lloyd's bound \cite{LloydBound}, inspecting the UV divergences that arise in the bulk computation \cite{Reynolds:2016rvl,Kim:2017lrw}, the inclusion of quantum bulk effects \cite{Emparan:2021hyr}, noncommutative gauge theory \cite{Couch:2017yil}, the effects of the presence of a conformal anomaly \cite{Avila:2020ved}, and many other works. 

However, several aspects of the holographic complexity remain ambiguous. One of said aspects is the choice of the reference state $|R\rangle$ in the definition of the CV conjecture \eqref{CVConjecture}. It would be natural to consider the vacuum $|0\rangle$ as reference state, but direct bulk computations show that in general $\mathcal{C}(|0\rangle)$ is non-vanishing, so this is not such an obvious candidate. Nonetheless, it is possible to get a measure of the complexity of creating a given state $|\psi\rangle$ from the the vacuum $|0\rangle$ by computing the so-called complexity of formation, which is a vacuum-subtracted version of the complexity
\begin{equation}
\mathcal{C}_{F}(|\psi\rangle)=\mathcal{C}(|\psi\rangle)-\mathcal{C}(|0\rangle),
\label{CFDef}
\end{equation}
when both states are defined at $\tau=0$.

The holographic complexity of formation was first studied in \cite{Chapman:2016hwi}. In that work the authors investigated what is the additional complexity involved in forming an entangled thermofield double state (TFD) 
\begin{equation}
|TFD\rangle=\frac{1}{Z^{\frac{1}{2}}}\sum_{n}e^{-\frac{\beta E_{n}}{2}}e^{-E_{n}(t_{L}+t_{R})}|E_{n}\rangle_{L}|E_{n}\rangle_{R},
\end{equation}
compared to preparing each of the two individual CFT's in their vacuum state. According to the holographic dictionary \cite{Maldacena:2001kr}, the bulk dual to a TFD state is a two-sided eternal black hole. The authors of \cite{Chapman:2016hwi} compute the holographic complexity of AdS black holes in different dimensions and with different horizon topologies using both the CV and CA prescriptions. In all the cases studied there, $\mathcal{C}_{F}$ turned out to be positive or zero, never negative. 

In fact, one could argue that the complexity of formation needs to satisfy
\begin{equation}
\mathcal{C}_{F}\geq 0,
\label{CFPositive}
\end{equation}
as the vacuum should be the ``simplest'' state on any theory, with the equality occurring only if $|\psi\rangle=|0\rangle$. The positivity of $\mathcal{C}_{F}$ was investigated in particular spacetimes \cite{Fu:2018kcp,Chapman:2018bqj} and perturbations thereof \cite{Flory:2018akz,Bernamonti:2020bcf}, and it was proven to be true in  \cite{Engelhardt:2021mju} for the CV conjecture if the bulk geometry meets certain conditions. More specifically, if the state $|\psi\rangle$ is dual to an asymptotically $AdS_{d+1}$ spacetime and the latter satisfies the Weak Curvature Condition (WCC) 
\begin{equation}
t^{\mu}t^{\nu}\left(R_{\mu\nu}-\frac{1}{2}g_{\mu\nu}R-\frac{d(d-1)}{L^{2}}g_{\mu\nu}\right)\geq 0, \,\, \forall \, \text{timelike}\,\, t^{\mu},
\label{WCC} 
\end{equation}
which in Einstein gravity is equivalent to the Weak Energy Condition (WEC)
\begin{equation}
t^{\mu}t^{\nu}T_{\mu\nu}\geq 0, \,\, \forall \, \text{timelike}\,\, t^{\mu},
\label{WEC}
\end{equation}
with $T_{\mu\nu}$ the stress energy tensor of the theory, then the vacuum is the least CV complex state. The WCC requirement is essential, as the examples found in \cite{Chapman:2018bqj,Bernamonti:2020bcf} that give $\mathcal{C}_{F}<0$ also violate it. However, it was later shown in \cite{Engelhardt:2021kyp} that any solution to type IIB and eleven-dimensional SUGRA satisfies the WCC. In other words, even if the dimensional reduction of a specific asymptotically $AdS_{d+1}\times K$ spacetime over the compact manifold $K$ violates the WCC, the inclusion of this directions restores it. While at first glance this may suggest that the positivity of $\mathcal{C}_{F}$ always should hold from the higher dimensional point of view, this is not the case.

The first point to consider is that there is not an obvious and unique way to extend the CV prescription to include the compact directions. We have at the very least two natural generalizations: 
\begin{enumerate}
\item The volume of the maximal hypersurface $\Sigma_{full}$ in the full $AdS_{d+1}\times K$ spacetime
\item The volume of the hypersurface $\Sigma_{up}$ defined in the full $AdS_{d+1}\times K$ spacetime as the uplift of the hypersurface $\Sigma$ which has maximal volume in the asymptotically $AdS_{d+1}$ part of the bulk.
\end{enumerate}
Note that in general $\Sigma_{full}\neq \Sigma_{up}$, as the uplifted hypersurface that was maximal in $AdS_{d+1}$ does not need to still be maximal in $AdS_{d+1}\times K$. Hence in general
\begin{equation}
\text{Vol}(\Sigma_{full})\neq\text{Vol}(\Sigma_{up}),
\end{equation}
and these generalizations can yield different results. The positivity of the complexity of formation coming from this two candidates was studied in \cite{Engelhardt:2021kyp} in a special scenario where they coincide, and in which they conclude that neither can reliably avoid negative values of $\mathcal{C}_{F}$ even when the WCC is satisfied from the high dimensional point of view. The reason is that the inclusion of the compact directions violates the assumption of asymptotically $AdS_{d+1}$ boundary conditions used in the proof given in \cite{Engelhardt:2021mju}. The gravitational asymptotically $AdS_{4}\times S^{7}$ backgrounds considered in \cite{Engelhardt:2021kyp} are all part of a consistent truncation of eleven dimensional SUGRA over the $S^{7}$. The truncated theory features a tachyonic scalar field with mass above the Breitenlohner-Freedman (BF) bound, which causes the violation of the WCC (equivalent to the WEC in this case) from the lower dimensional point of view. The complexity of formation coming from both the truncated theory and the full eleven-dimensional one turn out to be negative.

Motivated by the findings in \cite{Engelhardt:2021kyp}, here we study the positivity of the complexity of formation from both the lower and higher dimensional points of view in two particular gravitational models. We do this by analysing the vacuum subtracted complexity when the states are defined not only at $\tau=0$, but at any finite time, with a quantity that we term `evolving complexity'. Both gravitational models come from the consistent truncation anzats of SUGRA IIB solutions given in \cite{Cvetic:1999xp}, making both optimal candidates to include the compact directions in the computation of the holographic complexity. The first of this is the D'Hoker and Kraus \cite{DHoker:2009mmn} model, which we will refer as DK for short, dual to finite temperature SYM $\mathcal{N}=4$ in the presence of an external magnetic field. The second one is the \'Avila and Pati\~no \cite{Avila:2018hsi} model, which we will refer as AP for short, also dual to finite temperature SYM $\mathcal{N}=4$ in the presence of an external magnetic field, but with the addition of a non zero vacuum expectation value (VEV) for a single trace scalar operator of scaling dimension equal to 2. An important feature of the AP model is that, at any given value for the source of the scalar operator, there exists a maximum magnetic field intensity $B_{c}$ that the plasma can tolerate, becoming unstable for higher values. Below $B_{c}$ there are two branches of solutions for any given magnetic field intensity, with one of them thermodynamically preferred over the other.

We will show below that in the DK model, the uplift of the hypersurface in the five dimensional truncation is also extremal in the full ten dimensional background, and therefore the two manners described above to include the compact directions are equivalent, and furthermore, the result of applying the CV prescription in either the five or ten dimensional theories is the same up to a constant factor of no consequence. On the contrary, in the AP model, the uplift of the extremal hypersurface in five dimensions is not only not extremal in ten dimensions, giving place to the discussion about how to incorporate the compact space, but also the complexity obtained using the uplift in ten dimensions is not simply proportional to the one extracted from the five dimensional truncation, providing no argument to prefer this strategy over the other.

Our results show that while in the case of the DK model the complexity of the state increases with the magnetic field, which is consistent with the intuition that it is harder to create a state with a finite $B$ compared to one with a vanishing $B$, in the case of the AP model the story is not so simple. As it could be arguably expected, the states in the thermodynamically unstable branch are less complex than the one without a magnetic field (but identical in every other aspect to any state of the branch). Surprisingly though, there are states in the stable branch that  are less complex than the $B=0$ state, occurring at a range of magnetic field intensities $B_{s}<B<B_{c}$ close to maximum that the background can bare. We call this phenomenon `magnetic simplification', and in order to study it we introduce a vanishing magnetic field subtracted version of the complexity, which we call `complexity of magnetization', defined as the complexity associated to magnetizing a given state. Given that the main difference between the DK and AP models is the inclusion of the scalar operator with a non vanishing VEV on the later, this leads us to conclude that the direct impact of the magnetic field is to increase the complexity of a state, but it can indirectly lower it by diminishing the complexity associated to additional degrees of freedom when these do not vanish across the space.

The manuscript is organized as follows. In Sec. \ref{GravitySetup} we review the construction of the DK and AP models, explaining how both are part of the same general truncation anzats. In Sec. \ref{Complexity_5D_Sec} we show how to compute the complexity by means of the CV prescription for both models from the five-dimensional perspective and present the numerical results, while in Sec. \ref{Complexity_10D} we do the same from the ten-dimensional point of view. We close by discussing our results in Sec. \ref{Discussion}. Some of the more technical details of our computations are contained in a series of appendices.
\section{Gravity setup}
\label{GravitySetup}
\subsection{General truncation anzats}
\label{Truncation}
The family of solutions that we consider in this work are part of the general truncation anzats given in \cite{Cvetic:1999xp}. We consider solutions to ten-dimensional SUGRA IIB in which the metric and the self-dual five-form are the only fields that are turned on. Upon reduction, the five-dimensional fields are the the metric $g_{\mu\nu}$, three Maxwell fields $A^{i}$ and two scalar fields $\varphi_{j}$. The explicit form of the self-dual five-form is irrelevant for the present work, while the ten-dimensional line element is given by
\begin{equation}
ds_{10}^{2}=\Delta^{\frac{1}{2}}ds_{5}^{2}+\frac{L^{2}}{\Delta^{\frac{1}{2}}}\sum_{i=1}^{3}X_{i}^{-1}\left(d\mu_{i}^{2}+\mu_{i}^{2}\left(d\phi_{i}+\frac{A^{i}}{L}\right)^{2}\right),
\label{met_10}
\end{equation}
where $ds_{5}^{2}$ is the line element of the truncated theory, $L$ is a parameter with units of length that corresponds to the $AdS_{5}$ radius, the $\mu_i$ coefficients are given by
\begin{equation}
\mu_{1}=\sin\theta, \qquad \mu_{2}=\cos\theta\sin\psi, \qquad \mu_{3}=\cos\theta\cos\psi,
\end{equation}
the wrapping factor $\Delta$ by
\begin{equation}
\Delta=\sum_{i=1}^{3}X_{i}\mu_{i}^{2},
\end{equation}
with
\begin{equation}
X_{i}=e^{-\frac{1}{2}\vec{a}_{i}\cdot\vec{\varphi}}, \qquad \vec{a}_{i}=(a_{i}^{(1)},a_{i}^{(2)}), \quad \vec{\varphi}=(\varphi_{1},\varphi_{2}),
\label{truncA}
\end{equation}
and the $\vec{a}_{i}$ must satisfy
\begin{equation}
\qquad \vec{a_{i}}\cdot\vec{a_{j}}=4\delta_{ij}-\frac{4}{3}.
\end{equation}
We are using Hopf (toroidal) coordinates \cite{LachiezeRey:2005hs,Achour:2015zpa} on the compact directions, which means that
\begin{equation}
0\leq\theta, \psi\leq\frac{\pi}{2}, \quad 0\leq \phi_{i}\leq 2\pi.
\end{equation}

Substitution of the reduction anzats in the SUGRA IIB equations of motion gives five-dimensional equations of motion that can be derived from the five dimensional effective action
\begin{equation}
\begin{split}
S&=\frac{1}{16\pi G_{5}}\int d^{5}x\sqrt{-g}\left(R-\frac{1}{2}\sum_{j=1}^{2}(\partial\varphi_{j})^{2}\right.\\&\left.+\frac{4}{L^{2}}\sum_{i=1}^{3}(X_{i}^{-1}-\frac{1}{4}X_{i}^{-2}(F^{i})^{2})+\frac{1}{4}\epsilon^{\mu\nu\rho\sigma\lambda}F^{1}_{\mu\nu}F^{2}_{\rho\sigma}A^{3}_{\lambda}\right),
\end{split}
\label{ReductionAction}
\end{equation}

Next we list the solutions to \eqref{ReductionAction} that we will study in this paper. In what follows we will set $L=1$ without loss of generality.
\subsection{Vacuum}
\label{AdS5S5}
The first solution is $AdS_{5}\times S^{5}$, which is dual to the vacuum state of SYM $\mathcal{N}=4$. As such, we will refer to it as the `vacuum solution'. In this background all the scalar and Maxwell fields are turned off
\begin{equation}
\varphi_{2}=\varphi_{1}=0, \qquad A^{1}=A^{2}=A^{3}=0,
\end{equation}
while the ten-dimensional line element is taken to be
\begin{equation}
ds_{10}^{2}=ds_{5}^{2}+d\theta^{2}+\sin^{2}\theta d\phi_{1}^{2}+\cos^{2}\theta d\Omega_{3}^{2}.
\label{dsAdS5S5}
\end{equation}
In Hopf coordinates the $S^{3}$ line element is written as
\begin{equation}
d\Omega_{3}^{2}=d\psi^{2}+\sin^{2}\psi d\phi_{2}^{2} +\cos^{2}\psi d\phi_{3}^{2},
\label{S3}
\end{equation}
while the Poincare $AdS_{5}$ line element is
\begin{equation}
ds^{2}_{5}=\frac{dr^{2}}{r^{2}}+r^{2}\left(-dt^{2}+dx^{2}+dy^{2}+dz^{2}\right).
\label{AdS5}
\end{equation}
We have chosen coordinates such that $(t,x,y,z)$ are the SYM theory directions and $r$ is the $AdS_{5}$ radial coordinate, with the boundary located at $r\rightarrow\infty$.
\subsection{Finite temperature}
\label{BB}
Another important solution is the Black D3-brane geometry, which is dual to SYM $\mathcal{N}=4$ at finite temperature $T$. Once again the scalar and Maxwell fields are turned off and the ten-dimensional line element is given by \eqref{dsAdS5S5} and \eqref{S3}. The difference is that the five-dimensional line element is now
\begin{equation}
ds^{2}_{5}=\frac{dr^{2}}{U_{BB}(r)}-U_{BB}(r)dt^{2}+\left(r+\frac{r_{h}}{2}\right)^{2}(dx^{2}+dy^{2}+dz^{2}),
\label{dsBB}
\end{equation}
where
\begin{equation}
U_{BB}(r)=\left(r+\frac{r_{h}}{2}\right)^{2}\left(1-\frac{\left(\frac{3}{2}r_{h}\right)^{4}}{\left(r+\frac{r_{h}}{2}\right)^{4}}\right).
\label{UBB}
\end{equation}
This geometry features a black hole whose event horizon is located at $r=r_{h}$, and its temperature, given by
\begin{equation}
T=U'(r)\vert_{r_h}=\frac{3 r_{h}}{2\pi},
\label{Temperature}
\end{equation}
dictates the one of the quantum state as well.

We are not using the standard $\tilde{r}$ radial coordinate and instead we employ a scaled and translated version $r$ defined by the relation $\tilde{r}=r+\frac{r_{h}}{2}$. The reason is that the next two families of solutions are naturally written in this coordinate. Of course, \eqref{dsBB} reduces to the vacuum solution \eqref{AdS5} when we take $T=0$.
\subsection{DK model}
\label{DKBackground}
Next is the family of solutions constructed by D'Hoker and Kraus \cite{DHoker:2009mmn}, which we will refer as the DK model for short. This model is dual to SYM $\mathcal{N}=4$ at a finite temperature $T$ in the presence of a magnetic field $B$. This can be recovered from the general truncation anzats by setting
\begin{equation}
\varphi_{2}=\varphi_{1}=0, \qquad A^{1}=A^{2}=A^{3}=2\frac{A}{\sqrt{3}},
\end{equation}
and thus the ten-dimensional line element takes the form
\begin{equation}
\begin{aligned}
ds_{10}^{2}=&ds_{5}^{2}+d\theta^{2}+\sin^{2}\theta\left(d\phi_{1}+\frac{2}{\sqrt{3}}A\right)^{2}\\&+\cos^{2}\theta d\sigma_{3}^{2}(A).
\end{aligned}
\label{10DDK}
\end{equation}
The presence of the Maxwell field $A$ deforms the 3-sphere in such a way that a 3-cycle with line element 
\begin{equation}
\begin{split}
d\sigma_{3}^{2}(A)=&d\psi^{2}+\sin^{2}\psi\left(d\phi_{2}+\frac{2}{\sqrt{3}}A\right)^{2}\\&+\cos^{2}\psi\left(d\phi_{3}+\frac{2}{\sqrt{3}}A\right)^{2},
\end{split}
\label{S3DK}
\end{equation}
is obtained.

On the other hand, the anzats for the line element of the non-compact part of the spacetime is
\begin{equation}
ds_{5}^{2}=\frac{dr^{2}}{U(r)}-U(r)dt^{2}+V(r)(dx^{2}+dy^{2})+W(r)dz^{2},
\end{equation}
while the only Maxwell field of the truncation is taken to be
\begin{equation}
F=B\,dx\wedge dy
\end{equation}
We are using the same coordinates as in the vacuum and black D3 solutions, and the construction is done so that like in those cases, every element of this latter family of backgrounds features a horizon located at $r_{h}$ where the metric function $U(r)$ vanishes. The metric asymptotes precisely $AdS_{5}$ at the boundary $r\rightarrow\infty$ for any $B$ and $T$, with the former matching the magnetic field intensity in the dual gauge theory. Thus every member of the family is characterized by the values of its magnetic field intensity $B$ and temperature $T$, which suggests labelling each solution by the dimensionless ratio $B/T^{2}$. However, the DK model features a conformal anomaly for any non-vanishing magnetic field intensity, which introduces another length scale at the quantum level in the gauge theory side. As a consequence not all dimensionless physical observables are functions of $B/T^{2}$ alone. We have previously shown in \cite{Avila:2021zhb} that, when computed by means of the CA conjecture, the holographic complexity is insensitive to the conformal anomaly in the DK model. We will investigate if this is also the case for the CV conjecture in the following sections.

The only known analytical members of the DK model are the black D3-brane solution for $B/T^{2}=0$ and BTZ$\times\mathbb{R}^{2}$ for precisely $B/T^{2}=\infty$. For any intermediate values of $B/T^{2}$ it is necessary to resort to numerical methods to solve the equations of motion. The explicit integration procedure that we follow is explained in detail in \cite{Arean:2016het} for the solutions outside the event horizon and in \cite{Avila:2018sqf,Avila:2021zhb} for the solutions inside the horizon. 
\subsection{AP model}
\label{APBackground}
Finally we have the family of solutions constructed by \'Avila and Pati\~no \cite{Avila:2018hsi}, which we will refer to as the AP model for short. This background is also dual to SYM $\mathcal{N}=4$ at finite temperature $T$ in the presence of a magnetic field $B$, but with a non-vanishing VEV (which is a function of $B$ and $T$) for a single trace scalar operator. The model is obtained from the general truncation anzats by taking
\begin{equation}
\frac{2}{\sqrt{3}}\varphi_{2}=2\varphi_{1}=\varphi,\quad A^{1}=0, \quad A^{2}=A^{3}=\sqrt{2}A,
\label{particular1}
\end{equation}
with
\begin{equation}
\vec{a}_{1}=\left(\frac{2}{\sqrt{6}},\sqrt{2}\right),\,\vec{a}_{2}=\left(\frac{2}{\sqrt{6}},-\sqrt{2}\right),\, \vec{a}_{3}=\left(-\frac{4}{\sqrt{6}},0\right),
\label{particular2}
\end{equation}
which in turn means that
\begin{equation}
X=X_{2}=X_{3}=e^{\frac{1}{\sqrt{6}}\varphi}, \qquad X_{1}=X^{-2},
\end{equation}
and the wrapping factor is given by
\begin{equation}
\Delta=X^{-2}\sin^{2}\theta+X\cos^{2}\theta.
\label{wrapping}
\end{equation}

The ten-dimensional line element is given by
\begin{equation}
\begin{split}
ds_{10}^{2}=&\Delta^{\frac{1}{2}}ds_{5}^{2}+\frac{1}{\Delta^{\frac{1}{2}}}\left(X\Delta d\theta^{2}+X^{2}\sin^{2}\theta d\phi_{1}^{2}\right. \\& \left.+X^{-1}\cos^{2}\theta d\Sigma_{3}^{2}(A)\right),
\end{split}
\label{10DAP}
\end{equation}
where the 3-cycle line element $d\Sigma_{3}^{2}(A)$ depends on the one Maxwell field of the truncation and is given by
\begin{equation}
\begin{split}
d\Sigma_{3}^{2}(A)=&d\psi^{2}+\sin^{2}\psi\left(d\phi_{2}+\sqrt{2}A\right)^{2}\\& +\cos^{2}\psi\left(d\phi_{3}+\sqrt{2}A\right)^{2}.
\end{split}
\label{S3AP}
\end{equation}
On the other hand, the anzats for the line element of the non-compact part of the spacetime is once again
\begin{equation}
ds_{5}^{2}=\frac{dr^{2}}{U(r)}-U(r)dt^{2}+V(r)(dx^{2}+dy^{2})+W(r)dz^{2},
\end{equation}
while the Maxwell field is taken to be
\begin{equation}
F=B\,dx\wedge dy,
\end{equation}
and the only scalar field of the truncation depends solely on the radial coordinate
\begin{equation}
\varphi=\varphi(r).
\end{equation}

Every element of the family features a black hole, with a horizon located at $r=r_{h}$ where the metric function $U(r)$ vanishes, and asymptotes $AdS_{5}$ at the boundary $r\rightarrow\infty$. Under this circumstances the magnetic field intensity $B$ coincides with the one in the dual gauge theory. Given that the equations of motion coming from \eqref{ReductionAction} are highly non-linear, their solution must be obtained numerically for any non-vanishing intensity of the magnetic field. The general integration procedure in the region outside the horizon is described in detail in \cite{Avila:2018hsi}, while for the inner region we describe it in App. \ref{AppInt}. Notably, the equations of motion require a non-constant scalar field $\varphi(r)$ for any non-vanishing magnetic field, which means that the DK model cannot be recovered from the AP for $B$ other than zero, in which case both reduce to the black D3-brane.

The near boundary behavior of the scalar field $\varphi$ is
\begin{equation}
\varphi\rightarrow\frac{1}{r^{2}}\left(\varphi_{0}+\psi_{0}\log{r}\right),
\end{equation}
which means that it saturates the BF bound \cite{Breitenlohner:1982jf,Bianchi:2001kw} and it is dual to a single trace scalar operator $\mathcal{O}_{\varphi}$ of scaling dimension equal to 2. According to the holographic dictionary, $\psi_{0}$ is dual to the source of the operator and $\varphi_{0}$ to its vacuum expectation value $\langle \mathcal{O}_{\varphi}\rangle$ \cite{Bianchi:2001kw}. From the gauge theory perspective, it makes sense to specify the source of the operator and then compute the vacuum expectation value that it generates in response to such source. 

It was shown in \cite{Avila:2018hsi} that for any given source $\psi_{0}$ there exists a critical magnetic field intensity $B_{c}$ that the plasma can tolerate, becoming unstable for higher values. From the dual gravitational perspective, beyond this critical value $B_{c}$, the geometries develop a naked singularity. Below $B_{c}$ there exist two branches of solutions for any given $B/T^{2}$ that differ in the value that $\langle \mathcal{O}_{\varphi}\rangle/T^{2}$ takes. One of these branches was exhibit to be thermodynamically preferred over the other \cite{Avila:2018hsi}, since the one with the higher value for $\langle \mathcal{O}_{\varphi}\rangle/T^{2}$ corresponds to a state with negative specific heat, higher free energy and lower entropy than the other, showing that the solutions with smaller $\langle \mathcal{O}_{\varphi}\rangle/T^{2}$ are thermodynamically preferred. Throughout this manuscript we will fix the source $\psi_{0}$ to 0, which means that the maximum magnetic field intensity that the background can bear is given by $B_{c}/T^{2}\simeq 11.24$.

The original motivation for the AP model was to find a feasible way to easily add fundamental degrees of freedom by means of the embedding of D7-branes in the probe limit. This objective was achieved in \cite{Avila:2019pua,Avila:2020ved}, where it was proven that the interplay between the magnetic and scalar fields leads to a very interesting thermodynamic behavior for the fundamental matter. The two properties of the metric associated to \eqref{10DAP} that permit an easy embedding of a D7-brane on it are that its components do not depend on the angular coordinate $\phi$, and that the direction that the latter coordinate represents remains orthogonal to the rest of the spacetime. The inclusion of the scalar field $\varphi$ was crucial for this to happen.

Finally, another important thing to note is that the AP model, just like the DK model, possesses a conformal anomaly for any $B\neq 0$. We will investigate if this has any effect on the CV computation in the following sections.
\section{Complexity 5D}
\label{Complexity_5D_Sec}
\subsection{CV computation}
\label{CV_computation}
\begin{figure}[ht!]
 \centering
 \includegraphics[width=0.40\textwidth]{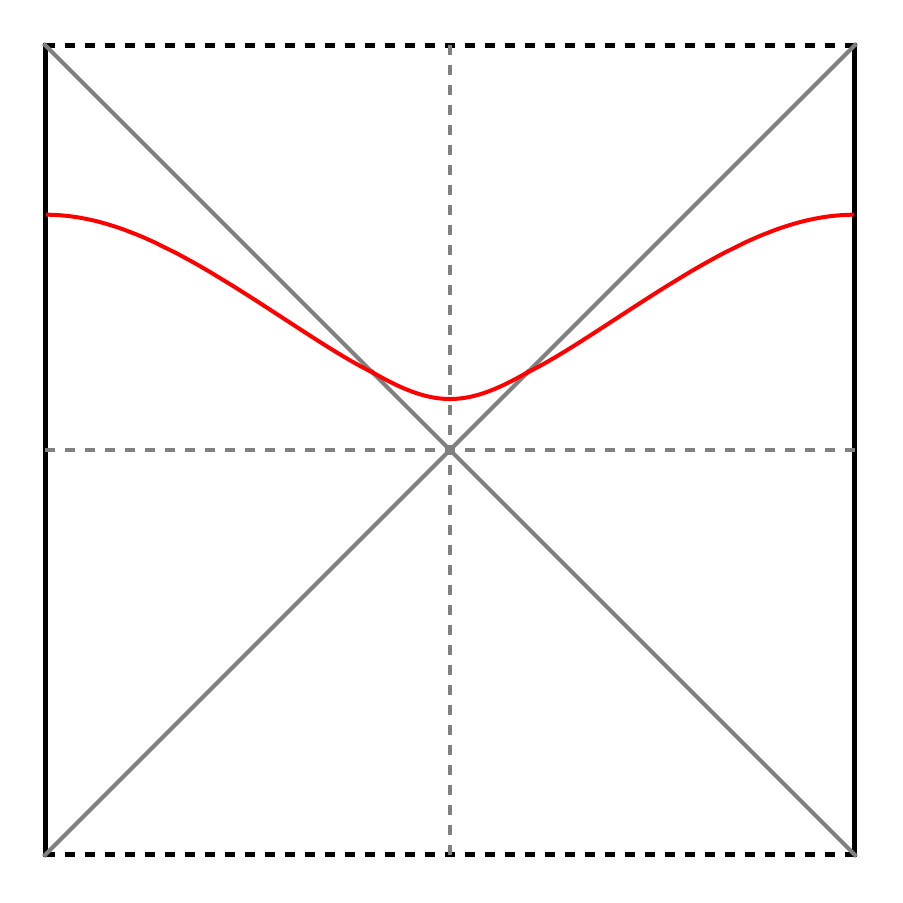}
\put(-5,155){$\tau$}
 \put(-205,155){$\tau$}
 \put(-105,120){$r_{m}$}
\caption{\small Penrose diagram for the class of geometries that we consider. The event horizon at $r=r_{h}$ is displayed as the gray continuous lines. The continuous red line denotes the codimension-one maximal hypersurface $\Sigma$ anchored at the time slice $t=\tau$ at both the left and right boundaries. There is a radius $r_{m}$ associated to any value of this $\tau$ given by the $r$ at which the embedding function $t(r)$ of the hypersurface satisfies $t(r_{m})=0$.}
\label{Penrose}
\end{figure}

In this section we will discuss how to compute the computational complexity for the two models described above when studied from the perspective of the truncated 5-dimensional theories. First we explain how to compute the complexity by means of the CV prescription in this class of bulk geometries. According to the CV conjecture the computational complexity $\mathcal{C}$ of a given gauge theory state $|\psi(\tau)\rangle$ at time $\tau$ is given by the volume of the maximal codimension-one hypersurface $\Sigma$ anchored at the time slice defined by $t=\tau$ at both the left and right boundaries. The concrete expression is
\begin{equation}
\mathcal{C}(|\psi(\tau)\rangle)=\text{max}_{\Sigma}\frac{\text{Vol}(\Sigma)}{G_{N}L},
\end{equation}
where $G_{N}$ is Newton constant and $L$ is an arbitrary length scale which we will take to be the AdS radius. In FIG. \ref{Penrose} we show an example of one of these hypersurfaces in the Penrose diagram for the class of geometries that we consider. The details of how to construct said Penrose diagram can be consulted in App. \ref{AppPenrose}.

The line element of every geometry in both the DK and AP model can be written as
\begin{equation}
ds_{5}^{2}=\frac{dr^{2}}{U(r)}-U(r)dt^{2}+V(r)(dx^{2}+dy^{2})+W(r)dz^{2}.
\label{5DBulk}
\end{equation}
In this coordinate system the desired codimension-one hypersurface $\Sigma$ can be parameterized as $x^{\mu}(\xi^{a})$, where $\mu$ runs across all the five directions of the bulk and $a$ runs across the four coordinates on the hypersurface. While this describes the most general embedding, the metric \eqref{5DBulk}, being diagonal with elements that depend on the radial coordinate alone, allows to chose the parametrization $\xi^{a}=(r, x, y, z)$ and $x^{\mu}(\xi^{a})=(t(r), r, x, y, z)$. With this choice the line element of the induced metric on the hypersurface $\Sigma$ is given by
\begin{equation}
\begin{split}
ds_{\Sigma}^{2}=&\left(\frac{1}{U(r)}-U(r)t'(r)^{2}\right)dr^{2}\\&+V(r)(dx^{2}+dy^{2})+W(r)dz^{2},
\end{split}
\end{equation}
where the prime denotes the derivative with respect to $r$, and the volume of $\Sigma$ can be computed as
\begin{equation}
\begin{split}
\text{Vol}(\Sigma)&=2V_{x}\int_{r_{m}}^{r_{\infty}}V(r)\sqrt{W(r)\left(\frac{1}{U(r)}-U(r)t'(r)^{2}\right)}dr \\&
=2V_{x}\int_{r_{m}}^{r_{\infty}}\mathcal{L}(t, t', r),
\end{split}
\label{5D_Vol1}
\end{equation}
where we have factorized the volume $V_{x}$ coming from the gauge theory spatial directions and the integration over $r$ runs from the minimal radius $r_{m}$ in the middle of the Penrose diagram (at $t=0$) to a regulator at $r_{\infty}$ near the boundary (hence the overall factor of 2). To obtain the precise result we will take the limit $r_{\infty}\rightarrow\infty$ at the end of the calculation.

According to the CV prescription we need to maximize the volume \eqref{5D_Vol1}. Extremization yields the following equation of motion for $t(r)$:
\begin{equation}
0=-\partial_{r}\left(\frac{\partial\mathcal{L}}{\partial t'}\right),
\label{5D_EOM1}
\end{equation}
which can be integrated to give
\begin{equation}
E=-\frac{\partial\mathcal{L}}{\partial t'}=-\frac{t'(r)U(r)V(r)\sqrt{W(r)}}{\sqrt{\frac{1}{U(r)}-U(r)t'(r)^{2}}},
\label{5D_EOM}
\end{equation}
or equivalently
\begin{align}
t'(r) = \frac{E}{U(r) \sqrt{E^2+U(r) V(r)^2 W(r)}},
\label{t_5D}
\end{align}
where $E$ is a conserved quantity. The  hypersurface $\Sigma$ needs to connect the boundary at the left with the one on the right without developing a conical singularity in the middle of the Penrose diagram at $r=r_{m}$ (see FIG. \ref{Penrose}). This is achieved by demanding that the derivative of $t(r)$ diverges at $r_{m}$, which by means of \eqref{5D_EOM} fixes the value of the constant $E$ to
\begin{equation}
E^{2}=-U(r_{m})W(r_{m})V(r_{m})^{2}.\label{ENS}
\end{equation}
For any given $r_{m}$ there is only one solution with the constant $E$ set by \eqref{ENS} that satisfies $t(r_{\infty})=\tau$ on both sides of the geometry, hence effectively we have $E=E(\tau)$ and $r_{m}=r_{m}(\tau)$. After substitution of \eqref{5D_EOM} in \eqref{5D_Vol1} we obtain the volume of the maximal hypersurface $\Sigma$ as a function of $\tau$
\begin{equation}
\text{Vol}(\Sigma)=2V_{x}\int_{r_{m}(\tau)}^{r_{\infty}}\frac{V(r)^{2}W(r)}{\sqrt{E(\tau)^{2}+U(r)V^{2}(r)W(r)}}dr,
\label{5D_Vol}
\end{equation}
where the limit $r_{\infty}\rightarrow\infty$ is meant to be taken.

In order to obtain the explicit dependence of Vol$(\Sigma)$ on $\tau$ we need to solve \eqref{5D_EOM} for $t(r)$. Given that, as explained in Sec. \ref{GravitySetup}, in general the backgrounds that are part of either the DK and AP models are constructed numerically, the solution for $t(r)$ for $\tau\neq 0$ needs to also be computed by numerical methods (with $t(r)=0$ for all $r$ being the only analytical solution). The integration procedure that we followed in practice began by solving \eqref{5D_EOM1} as a Frobenius expansions around $r_{m}$. Given that we look for solutions that satisfy $t(r_{m})=0$ and $t'(r_{m})=\infty$ the series turns out to be
\begin{equation}
t(r)=(r-r_{m})^{\frac{1}{2}}\sum_{i=0}^{\infty}t^{(m)}_{i}(r-r_{m})^{i},
\label{5D_Series_rm}
\end{equation}
where any coefficient $t^{(m)}_{i}$ can be determined using the equation of motion up to the necessary order. Of particular importance is the explicit expression for $t^{(m)}_{0}$
\begin{equation}
t^{(m)}_{0}=\left.2\frac{\sqrt{V W}}{\sqrt{-U(2U W V'+V(W U)')}}\right|_{r_{m}},
\end{equation}
because from it, and given that $U(r_{m})<0$, we can conclude that obtaining a real valued solution restricts $r_{m}$ to the interval $r_{min}<r_{m}\leq r_{h}$, where the minimal possible radius $r_{min}$ is given by the solution to the equation
\begin{equation}
0=(2U W V'+V(W U)')|_{r_{min}}.
\label{rmin}
\end{equation}
In the case of the DK and AP models, this minimal radius is a function of both the magnetic field intensity $B$ and the temperature $T$.

Once the coefficients $t^{(m)}_{i}$ are known to the desired order, Eq. \eqref{5D_Series_rm} can be used to provide initial conditions for the numerical integration of \eqref{5D_EOM1} starting at $r=r_{m}+\epsilon$, with $\epsilon\ll r_{m}$,  and only up to $r=r_{h}-\epsilon$, as the horizon is another singular point of the equation of motion. A series expansion of \eqref{5D_EOM1} near $r_{h}$ reveals that $t(r)$ goes like
\begin{equation}
t(r)=-\frac{1}{6r_{h}}\log(|r-r_{h}|)+\sum_{j=0}^{\infty}t^{(h)}_{j}(r-r_{h})^{j},
\label{5D_Series_rh}
\end{equation}
where any $t^{(h)}_{j}$ for $j\geq 2$ can be written in terms of $t^{(h)}_{0}$ and $t^{(h)}_{1}$. In practice we extracted $t^{(h)}_{0}$ and $t^{(h)}_{1}$ from the behavior of the numerical interior solution $t(r)$ near $r_h$, substitute these values in \eqref{5D_Series_rh}, and used the resulting series to provide initial conditions for the exterior numerical integration starting at $r=r_{h}+\epsilon$ and up to $r=r_{\infty}$. After mirroring this result for the left side of the Penrose diagram, this procedure allows us to piecewise construct the solution $t(r)$ for any $r$. Finally, we extracted $\tau$ from the numerical solution as $t(r_{\infty})=\tau$ and then obtained the relations $r_{m}(\tau)$ and $E(\tau)$.

\begin{figure}[ht!]
 \centering
 \includegraphics[width=0.20\textwidth]{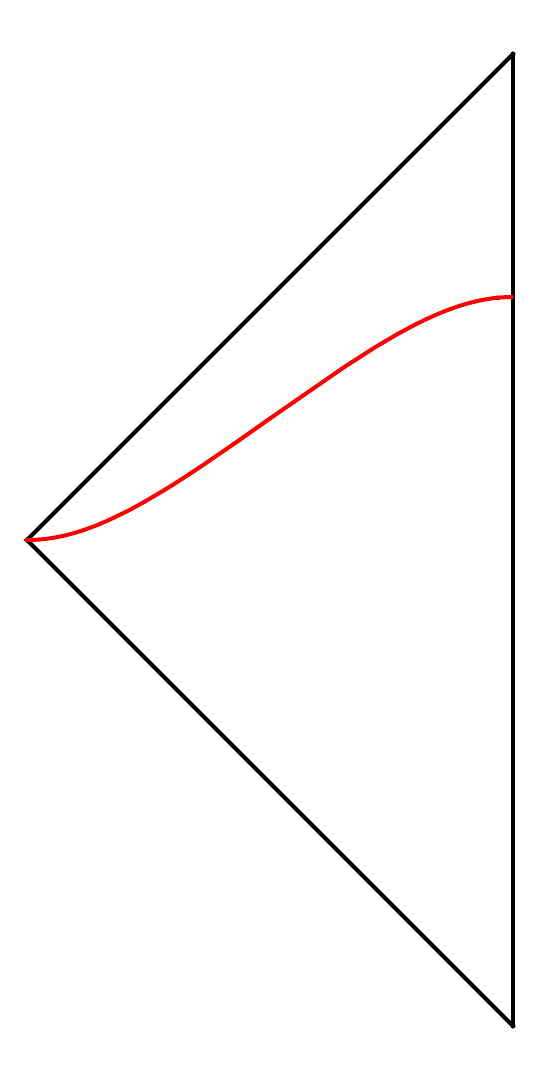}
\put(0,150){$\tau$}
\caption{\small Penrose diagram for one copy of the Poincare $AdS_{5}$ background dual to the vacuum state in both the DK and AP models. The continuous red line denotes the codimension-one maximal hypersurface $\Sigma$ anchored at the time slice $t=\tau$ at the boundary. For this background, $\Sigma$ is given by constant $t$ hypersurfaces.}
\label{Penrose_AdS}
\end{figure}

The computation for the vacuum state requires its own discussion, as the integration procedure we just described does not apply even if \eqref{5D_EOM} does. Obtaining the complexity of preparing both the left and right gauge theories in their vacuum state requires working with two separate copies of the Poincare $AdS_{5}$ bulk geometry, where the maximal hypersurfaces $\Sigma_{0}$ are those with constant time, given by $t(r)=\tau$ which is equivalent to setting $E=0$ for any $\tau$ in \eqref{5D_EOM}. This in turn implies that the maximal volume \eqref{5D_Vol} for the vacuum state is given by
\begin{equation}
\text{Vol}(\Sigma_{0})=2V_{x}\int_{0}^{r_{\infty}}r^{2}dr=\frac{2 V_{x}}{3}r^{3}_{\infty},
\label{VolVacuum}
\end{equation}
showing that the volume is independent of the boundary time $\tau$. In FIG. \ref{Penrose_AdS} we present one copy of the Poincare $AdS_{5}$ bulk geometry with an example of a maximal hypersurface $\Sigma_{0}$.

It is important to note that for any of the geometries in both the DK and AP models, $\text{Vol}(\Sigma)$ is divergent when the boundary regulator $r_{\infty}$ is removed. Substitution of the near boundary expansions of the metric fields given in App. \ref{AppBdry} for either the DK or AP models in \eqref{5D_Vol} gives
\begin{equation}
\text{Vol}(\Sigma)=2 V_{x}\left(\frac{r^{3}_{\infty}}{3}+\frac{U_{1}r_{\infty}^{2}}{2}+\frac{U_{1}^{2}r_{\infty}}{4}+\mathcal{O}\left(r^{-1}\right)\right),
\end{equation}
which diverges in the limit $r_{\infty}\rightarrow\infty$. Note however that the previous expression can be rewritten as
\begin{equation}
\text{Vol}(\Sigma)=2 V_{x}\left(\frac{\tilde{r}^{3}_{\infty}}{3}+\frac{U_{1}^{3}}{24}+\mathcal{O}\left(r^{-1}\right)\right),
\label{Divergences}
\end{equation}
with $\tilde{r}_{\infty}=r_{\infty}+U_{1}/2$. Given that when $r_{\infty}\rightarrow\infty$ we have $\tilde{r}_{\infty}=r_{\infty}$, formally using either regulator at the boundary will give the same result for $\text{Vol}(\Sigma)$ once the limit has been taken. Using $\tilde{r}_{\infty}$ in $\text{Vol}(\Sigma)$ and $r_{\infty}$ in $\text{Vol}(\Sigma_{0})$ explicitly shows that the vacuum subtracted volume $\text{Vol}(\Sigma)-\text{Vol}(\Sigma_{0})$ is finite in the limit $r_{\infty}\rightarrow\infty$, that is, when both regulators are removed. This mathematical trick is necessary because of the choice of radial coordinate for both the DK and AP models. 

\subsection{Results}
\label{results5D}
The numerical procedure detailed above allows us to use \eqref{5D_Vol} and \eqref{CFDef} to find the computational complexity $\mathcal{C}$ of any state in the gauge theory dual to either the DK or AP models as a function of the three independent gauge theory parameters $B$, $T$ and $\tau$ as
\begin{equation}
\mathcal{C}(B,T,\tau)=\frac{\text{Vol}(\Sigma)}{G_{N}}.
\end{equation}
However, our numerical results show that the dimensionless ratio $\mathcal{C}/T^{3}$ only depends on the dimensionless quantities $B/T^2$ and $T \tau$, in terms of which the results ahead will be reported. Although at first sight this might seem trivial, this is not the case because, as explained in Sec. \ref{Intro}, both models feature a conformal anomaly that introduces an arbitrary energy scale $\mu$ at the quantum level. In other words, our results explicitly show that the complexity computed by means of the CV prescription is insensitive to the conformal anomaly, at the very least for these specific models. We have previously shown in \cite{Avila:2021zhb} that this is also the case when using the CA prescription for the DK and the Mateos-Trancanelli anisotropic models.

\begin{figure}[ht!]
 \centering
 \includegraphics[width=0.4\textwidth]{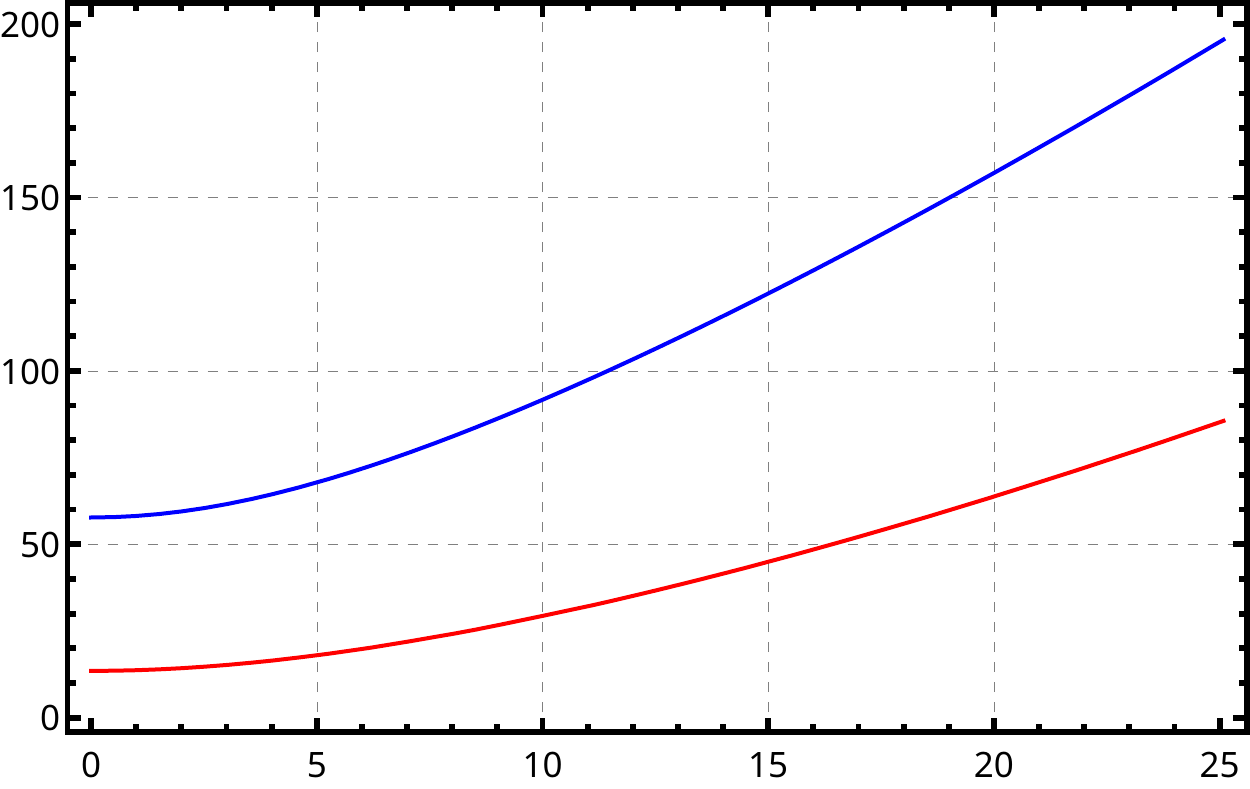}
\put(-225,110){\large $\frac{G_{N}\mathcal{C}_{E}}{V_{x}T^3}$}
\put(0,-10){\large $\frac{B}{T^2}$}
\caption{\small Evolving complexity $G_{N}\mathcal{C}_{E}/V_{x}T^{3}$ for the DK model as a function of $B/T^2$ with $T \tau =0$ (red, bottom) and $T \tau =1$ (blue, top). Both have a similar behavior, monotonically increasing as $B/T^2$ grows.}
\label{CF_DK_5}
\end{figure}

As previously explained, we are interested in the vacuum subtracted version of the complexity defined at any given $\tau$, a quantity that we call the evolving complexity $\mathcal{C}_{E}$, which is given by
\begin{equation}
\frac{\mathcal{C}_{E}(B/T^{2},T\tau)}{T^{3}}=\frac{\text{Vol}(\Sigma)-\text{Vol}(\Sigma_{0})}{T^{3}G_{N}},
\end{equation} 
Note that this quantity remains finite in the $r_{\infty}\rightarrow\infty$ limit by virtue of \eqref{Divergences}, and that it reduces to the well known complexity of formation for $\tau=0$. In FIG. \ref{CF_DK_5} we show $\mathcal{C}_{E}$ for the DK model as a function of $B/T^2$ for two different fixed values of $T \tau$. It can be seen that $\mathcal{C}_{E}$ is a monotonically increasing function of $B/T^2$ and that it is always positive $\mathcal{C}_{E}>0$ for the two values of $T\tau$ that we display. The interpretation of this result is that, at least intuitively, it becomes harder to create a state with a magnetic field starting from the vacuum as $B/T^2$ increases.

\begin{figure}[ht!]
 \centering
 \includegraphics[width=0.4\textwidth]{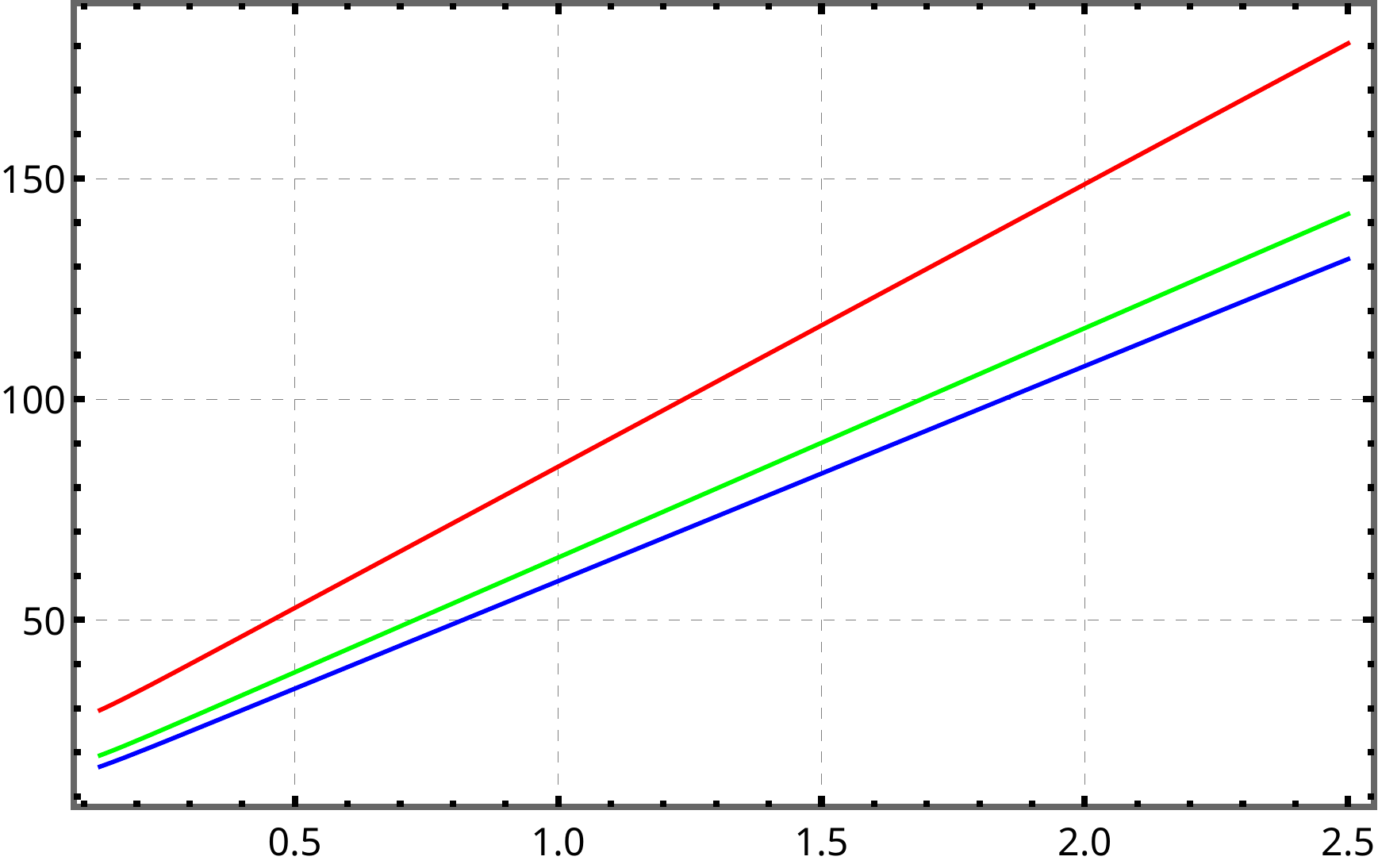}
\put(-225,120){\large $\frac{G_{N}\mathcal{C}_{E}}{V_{x}T^3}$}
\put(0,-10){$T\tau$}
\caption{\small Evolving complexity $G_{N}\mathcal{C}_{E}/V_{x}T^{3}$ for the DK model as a function of $T\tau$ for (bottom to top) $B/T^{2}= 0$ (blue), $B/T^{2}= 8.5$ (red) and $B/T^{2}= 18$ (green).}
\label{CF_DK_5_tau}
\end{figure}

From FIG. \ref{CF_DK_5} we can also see that the evolving complexity increases as the boundary time passes, as $\mathcal{C}_{E}$ is larger for $T\tau=1$ than it is for $T\tau=0$. This effect can be better appreciated in FIG. \ref{CF_DK_5_tau}, where we show $\mathcal{C}_{E}$ as a function of $T\tau$ for various values of $B/T^{2}$. It can be seen how $\mathcal{C}_{E}$ monotonically increases when $T \tau$ grows, in such a way that for late times it does it at a constant rate. With higher magnetic fields, the complexity grows even more, but keeps the same behavior, always increasing at a constant rate when $T \tau$ goes to infinity, which is the expected late time behavior of the computational complexity \cite{Susskind:2014rva,Stanford:2014jda,Carmi:2017jqz,Swingle:2017zcd,Mahapatra:2018gig,Alishahiha:2018tep,HosseiniMansoori:2018gdu,Auzzi:2022bfd,Avila:2021zhb}. While we arrived to this conclusion by inspecting the full time dependence of $C_{V}$, as a confirmation of our numerical procedure we present an alternative derivation of the $\tau\rightarrow\infty$ limit of $dC_{V}/d\tau$ explicitly App. \ref{dCdtau}

\begin{figure}[ht!]
 \centering
 \includegraphics[width=0.4\textwidth]{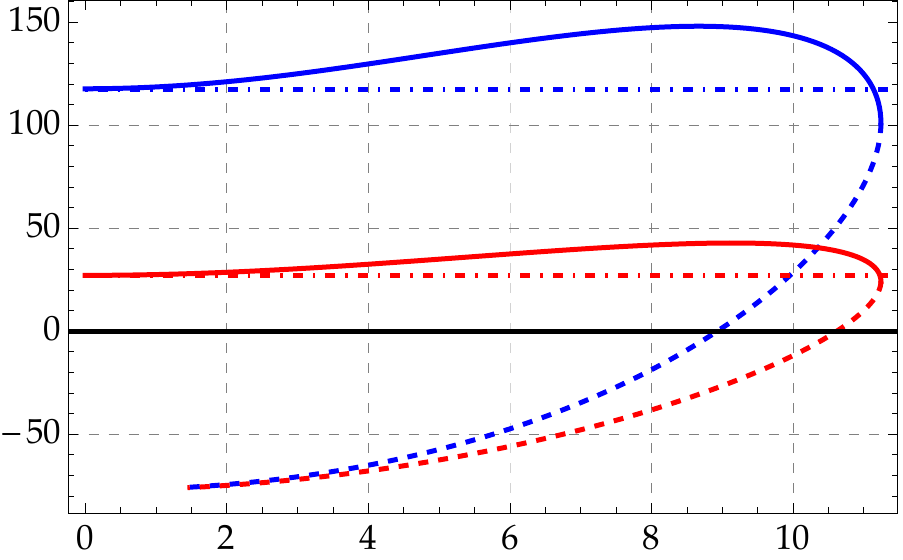}
\put(-225,110){\large $\frac{G_{N}\mathcal{C}_{E}}{V_{x}T^3}$}
\put(0,-10){\large $\frac{B}{T^2}$}
\caption{\small Evolving complexity $G_{N}\mathcal{C}_{E}/V_{x}T^{3}$ for the AP model as a function of $B/T^2$ with $T \tau =0$ (red, bottom) and $T \tau =1$ (blue, top). The solid lines represent states in the stable branch, while the dashed ones correspond to states in the unstable one. The horizontal dot-dashed lines denotes $\mathcal{C}_{E}$ for $B/T^{2}=0$ but the same $T\tau$ as the corresponding color.}
\label{CF_AP_5}
\end{figure}

In FIG. \ref{CF_AP_5} we show $\mathcal{C}_{E}$ as a function of $B/T^{2}$ at fixed values of $T\tau$ for the AP model.  As explained in Sec. \ref{Intro}, in the AP model two branches of solutions exists for any $0<B<B_{c}$, with one being thermodynamically preferred over the other. In the following plots we will denote the thermodynamically stable branch of states as a continuous line, while we will use a dashed line to indicate the latter. From FIG. \ref{CF_AP_5} it can immediately be seen that $\mathcal{C}_{E}$ is not a monotonic function of the magnetic field. As we increase the dimensionless quantity $B/T^{2}$ , both,  the complexity of formation ($T\tau=0$) and its evolution at $T\tau=1$ grow until they reach a maximum value. Interestingly, this maximum occurs for a magnetic field intensity lower than the critical one $B_{c}/T^{2}\approx 11.24$ for the two values of $T\tau$ displayed. Further increasing the magnetic field intensity causes $\mathcal{C}_{E}$ to decrease in such a way that there are some states that satisfy $C_{E}(B,T,\tau)<C_{E}(0,T,\tau)$ still within the stable branch, which can be stated as the system ongoing a `simplification' of sorts. In contrast to the DK model, in the AP model it is easier to create a state with a very intense magnetic field starting from the vacuum than it is to create one with a less intense magnetic field.

\begin{figure}[ht!]
 \centering
 \includegraphics[width=0.4\textwidth]{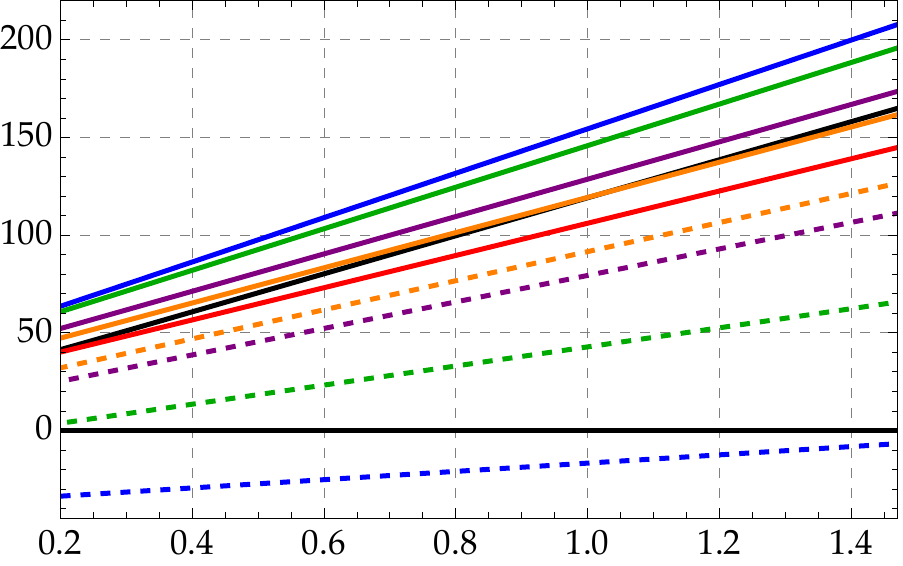}
\put(-225,105){\large $\frac{G_{N}\mathcal{C}_{E}}{V_{x}T^3}$}
\put(0,-10){$T \tau$}
\caption{\small Evolving complexity $G_{N}\mathcal{C}_{E}/V_{x}T^3$ for the AP model as a function of $T \tau$ with $B/T^{2} =  0$ as reference (black, forth line down),and then (from top to bottom for solid lines and bottom to top for dashed ones), $B/T^{2} \approx  8.08$ (blue), $10.37$ (green), $11.04$ (purple), $11.18$ (orange) and $11.24$ (red). The solid lines represent states in the stable branch, while the dashed ones correspond to states in the unstable one.}
\label{CF_AP_5_tau}
\end{figure}

In FIG. \ref{CF_AP_5_tau} we show $\mathcal{C}_{E}$ as a function of $T\tau$ for various $B/T^{2}$. First of all it can be seen that, just like in the case of the DK model, the evolving complexity monotonically increases with $T\tau$, in such a way that for late times it does at a constant rate for all the intensities of the magnetic field used in the plots, which is the expected behavior. Second, we confirm that indeed some of the states in the stable branch are such that for some $T\tau$ satisfy $C_{E}(B/T^{2},T\tau)<C_{E}(0,T\tau)$ as, for example, the orange continuous curve corresponding to the stable state with $B/T^{2}=11.18$ is below the black curve corresponding to $B/T^{2}=0$ for late $T\tau$.

This puzzling behavior rises the following question: is this simplification effect caused by the presence of the magnetic field alone or can it be attributed to the interplay that it has with the scalar field? In order to answer this we would like to subtract the contribution coming from the temperature from the complexity. We call this quantity the `complexity of magnetization' $\mathcal{C}_M$ of the state $|B,T,\tau\rangle$, defined as
\begin{equation}
\mathcal{C}_M(|B,T,\tau\rangle):= \mathcal{C}(|B,T,\tau\rangle) - \mathcal{C}(|0,T,\tau\rangle).
\end{equation}
Intuitively, $\mathcal{C}_{M}$ measures how difficult it is to prepare a state with a certain magnetic field and temperature at time $\tau$ starting from a state with the same temperature and at the same time, but with no magnetic field.

\begin{figure}[ht!]
 \centering
 \includegraphics[width=0.4\textwidth]{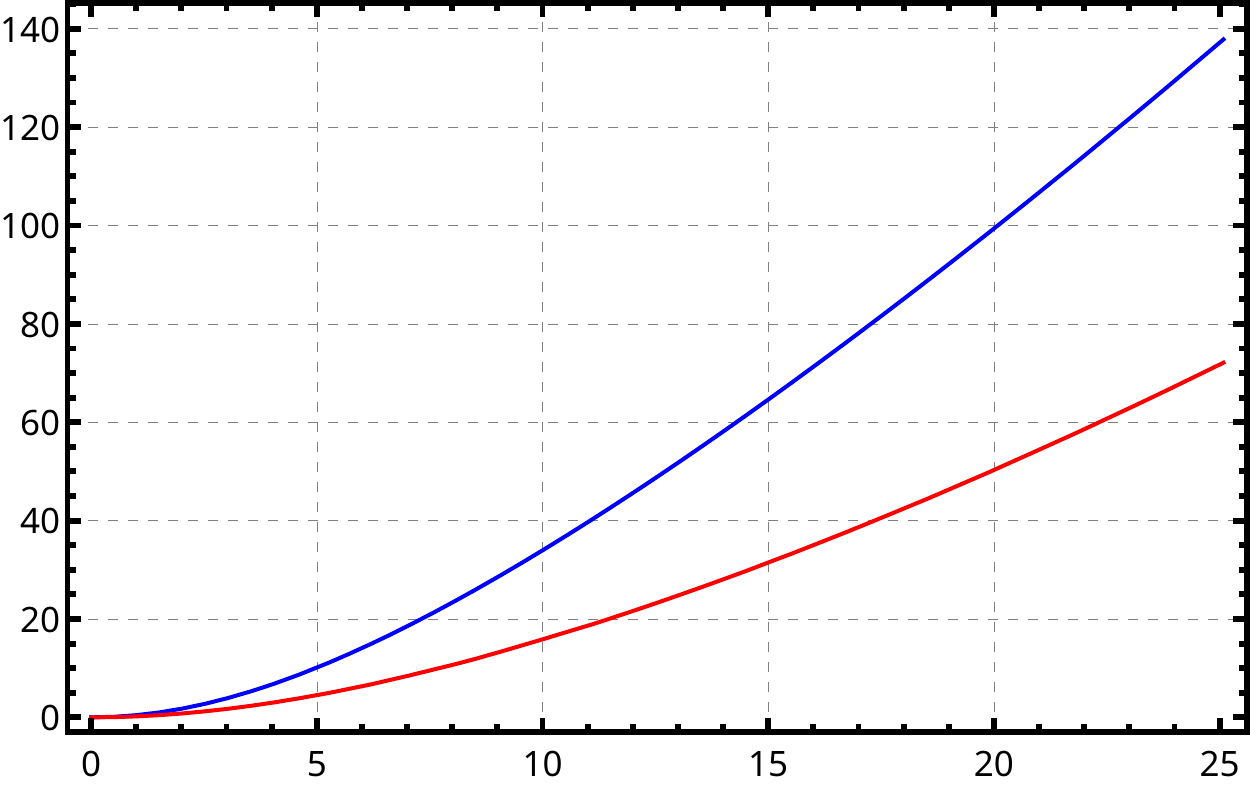}
\put(-230,110){\large $\frac{G_{N}\mathcal{C}_{M}}{V_{x}T^3}$}
\put(0,-10){\large $\frac{B}{T^2}$}
\caption{\small Complexity of magnetization $G_{N}\mathcal{C}_{M}/V_{x}T^3$ for the DK model as a function of $B/T^{2}$ for fixed $T \tau =0$ (red, bottom) and $T \tau =1$ (blue, top). Both have a similar behavior, monotonically increasing as $B/T^2$ grows.}
\label{CM_DK_5}
\end{figure}

In terms of the CV prescription, the gravity formula for $\mathcal{C}_{M}$ is
\begin{equation}
\frac{\mathcal{C}_{M}(B/T^{2},T\tau)}{T^{3}}=\frac{\text{Vol}(\Sigma)-\text{Vol}(\Sigma_{T})}{T^{3}G_{N}},
\end{equation} 
where $\Sigma_{T}$ is the maximal hypersurface anchored at fixed boundary time $T \tau$ in the Black D3-brane background, which corresponds to the $B/T^{2}=0$ solution for both the DK and AP models. Note that, just like the evolving complexity, $\mathcal{C}_M$ is finite in the limit $r_{\infty}\rightarrow\infty$ by virtue of \eqref{Divergences}.

\begin{figure}[ht!]
 \centering
 \includegraphics[width=0.4\textwidth]{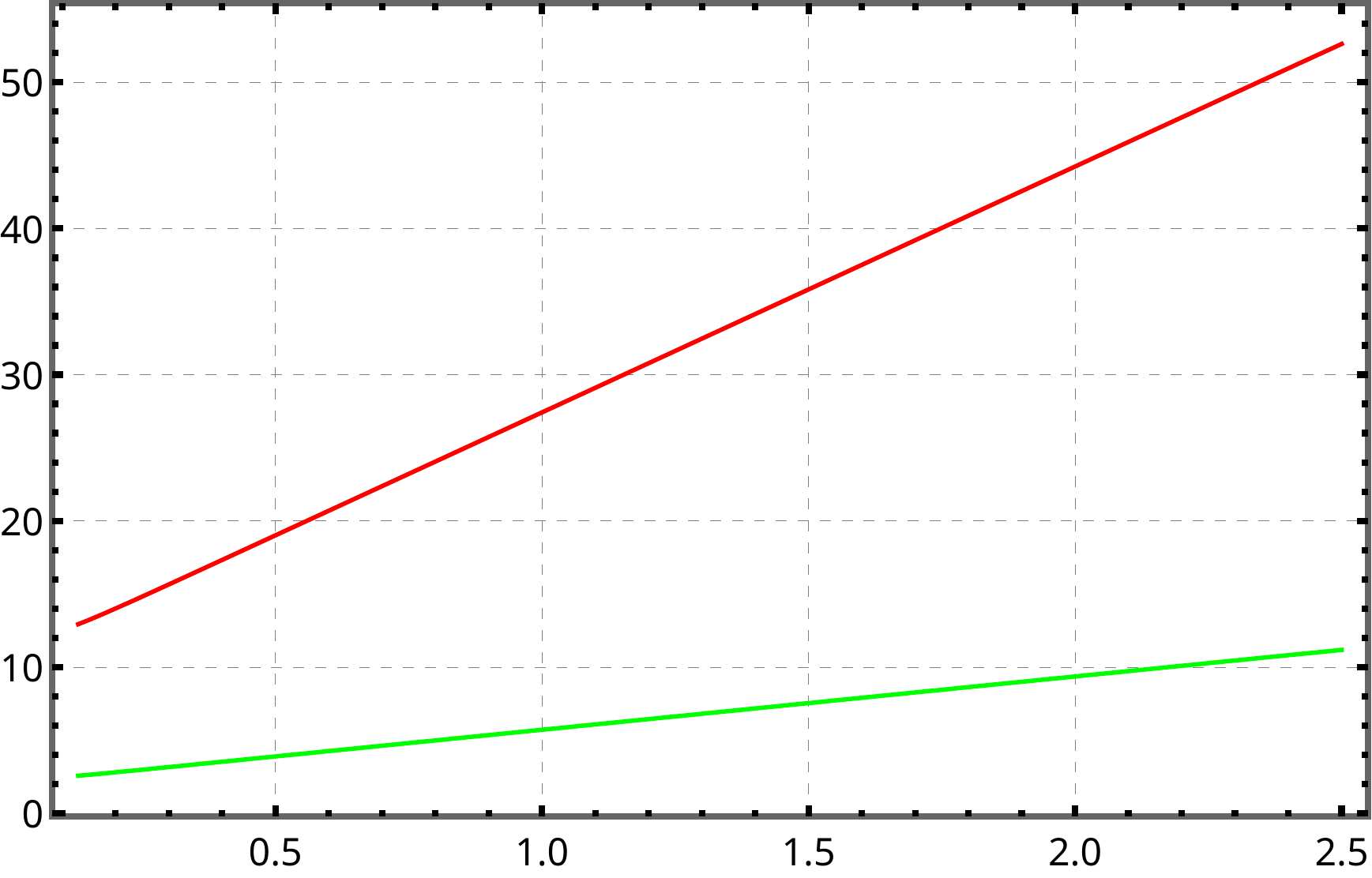}
\put(-230,105){\large $\frac{G_{N}\mathcal{C}_{M}}{V_{x}T^3}$}
\put(0,-10){$T\tau$}
\caption{\small Complexity of magnetization $G_{N}\mathcal{C}_{M}/V_{x}T^3$ for the DK model as a function of $T \tau$ for $B/T^{2}= 8.5$ (red, bottom) and $B/T^{2}= 18$ (green, top).}
\label{CM_DK_5_tau}
\end{figure}

We show the complexity of magnetization for the DK model as a function of $B/T^{2}$ at fixed $T\tau$ in FIG. \ref{CM_DK_5}, and as a function of $T\tau$ at fixed $B/T^{2}$ in FIG. \ref{CM_DK_5_tau}. From the first one, we can see that $\mathcal{C}_{M}$ is a monotonically increasing function of the magnetic field intensity for fixed $T \tau$ and that it is a positive quantity for all the explored values of $B/T^{2}$. From the latter we can see  a similar behavior, meaning that the complexity of magnetization always increases as $T \tau$ grows. Also note that, for the magnetic field intensities displayed in FIG. \ref{CM_DK_5_tau}, $C_M$ remains positive as $T \tau$ grows and that it increases at a constant rate as $T \tau$ goes to infinity.

\begin{figure}[ht!]
 \centering
 \includegraphics[width=0.4\textwidth]{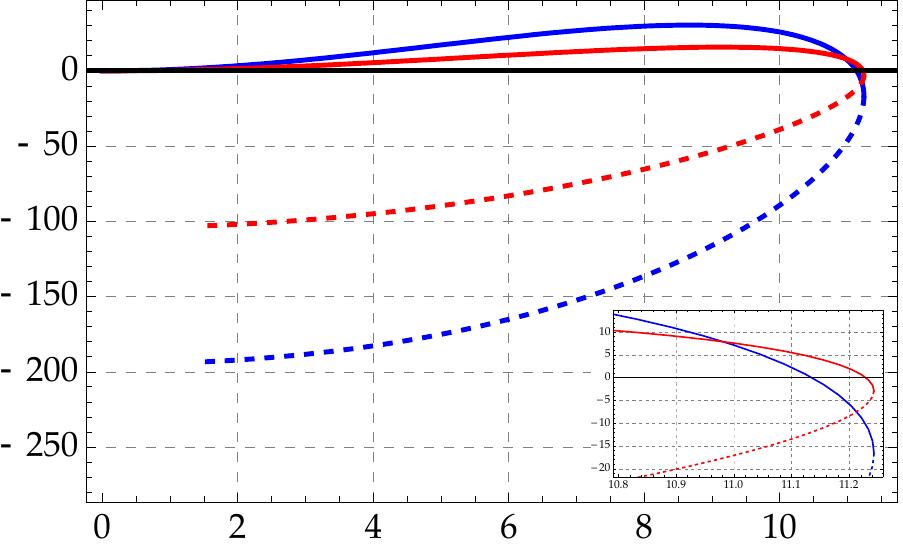}
\put(-220,105){\large $\frac{G_{N}\mathcal{C}_{M}}{V_{x}T^3}$}
\put(0,-10){\large $\frac{B}{T^2}$}
\caption{\small Complexity of magnetization $G_{N}\mathcal{C}_{M}/V_{x}T^3$ for the AP model as a function of $B/T^2$ with $T \tau =0$ (red, closer to zero) and $T \tau =1$ (blue, farther from zero). The solid lines represent states in the stable branch, while the dashed ones correspond to states in the unstable one. We present an inset of the large $B/T^{2}$ region in order to better visualize the magnetic simplification phenomenon.}
\label{CM_AP_5}
\end{figure}

In the case of the AP model, the complexity of magnetization reveals a more interesting behavior. In FIG. \ref{CM_AP_5} we show $\mathcal{C}_{M}$ as a function of $B/T^{2}$ for two values of $T \tau$, from which we can see that in both cases every state in the unstable branch is less complex than the thermal $B=0$ state, as for these we have that $\mathcal{C}_{M}<0$. However, as anticipated from the previous analysis of the evolving complexity, some of the states on the thermodynamically preferred branch also satisfy $\mathcal{C}_{M}<0$.

\begin{figure}[ht!]
 \centering
 \includegraphics[width=0.4\textwidth]{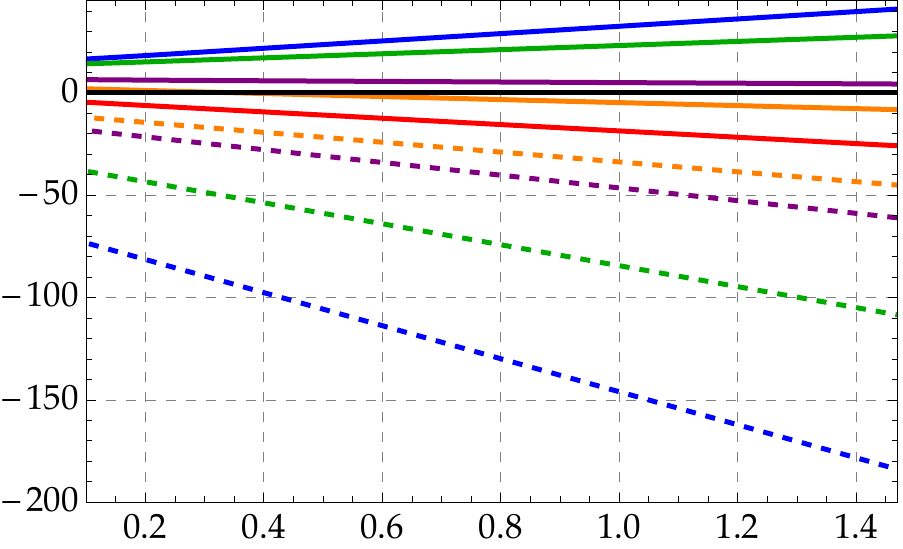}
\put(-220,105){\large $\frac{G_{N}\mathcal{C}_{M}}{V_{x}T^3}$}
\put(0,-10){$T \tau$}
\caption{\small Complexity of magnetization $G_{N}\mathcal{C}_{M}/V_{x}T^3$ for the AP model as a function of $T \tau$ with $B/T^2 =  0$ (black, at zero) as a reference, and then (from top to bottom for solid lines and bottom to top for dashed ones) $B/T^2 \approx  8.08$ (blue), $B/T^2 \approx  10.37$ (green), $B/T^2 \approx  11.04$ (purple), $B/T^2 \approx  11.18$ (orange) and $B/T^2 \approx  11.24$ (red). The solid lines represent states in the stable branch, while the dashed ones correspond to states in the unstable one.}
\label{CM_AP_5_tau}
\end{figure}

While the previous behavior is shown explicitly for the two values of $T\tau$ considered in FIG. \ref{CM_AP_5}, we can see that it is shared for other boundary times as well. In FIG. \ref{CM_AP_5_tau} we show the complexity of magnetization as a function of $T\tau$ for various values of the magnetic field. From this it can be seen that the states on the unstable branch have negative $\mathcal{C}_{M}$ for any $T\tau$, and that the complexity of magnetization grows at a constant rate as $T\tau$ increases. Notably the same is true for some of the states in the stable branch. For example, the continuous orange (bottom) curve in FIG. \ref{CM_AP_5_tau} corresponding to the stable state at $B/T^{2}=11.18$ satisfies $\mathcal{C}_{M}<0$ for $T\tau>0.6$. 

From the previous discussion we can conclude that indeed the interplay between the magnetic field and the scalar field leads to a negative complexity of magnetization $\mathcal{C}_{M}$, a phenomenon that we call `magnetic simplification'. This occurs for states with a magnetic field intensity such that  $B_{s}/T^{2} < B/ T^2 < B_{c}/ T^2$, where the simplification intensity $B_{s}$ depends on the time $T\tau$ at which we are defining the state.
 
\section{Complexity 10D}
\label{Complexity_10D}

As we previously mentioned, there is no obvious generalization of the CV prescription to allow the inclusion of the compact directions in ten dimensions. However, two natural options are: (1) uplift the maximal volume $\Sigma$ in 5D to $\Sigma_{up}$ in 10D and (2) find the maximal volume slice in the full 10D geometry $\Sigma_{full}$.

The volume of the hypersurface $\Sigma_{10}$ can be computed in a similar way to the five dimensional case. Now the coordinate system for the codimension-one hypersurface will be parameterized as $x^{\mu}(\xi^{a})$ where $\mu$ runs across the full ten dimensions of the bulk and $a$ across the nine directions of the hypersurface. We will again use the symmetries of the system to simplify the embedding.

We consider the DK model first and notice that the first term on the right side of \eqref{10DDK} depends solely on r, while in the others only the compact directions appear. This separation allows us to choose the parametrization $\xi^{a} = (r,x,y,z,\theta, \psi,\phi_1,\phi_2, \phi_3)$ and $x^{\mu}(\xi^{a}) = (t(r),r,x,y,z,\theta, \psi, \phi_{1},\phi_{2}, \phi_{3})$ in terms of which the volume of $\Sigma_{10}$ is
\begin{align}
\text{Vol}(\Sigma_{10}) = 2 \int \mathcal{L}_{10} d^9x,
\end{align}
where $\mathcal{L}_{10}$ is given by
\begin{align}
\mathcal{L}_{10} = \Lambda V(r) \sqrt{W(r)}  \sqrt{\frac{1}{U(r)}-U(r) t'(r)^2},
\end{align}
with
\begin{align}
\Lambda = \sin (\theta ) \cos ^3(\theta ) \sin (\psi ) \cos (\psi ).
\end{align}
It is because of this factorization of the compact directions that we can integrate them immediately, leading to a $\pi^3$ constant factor that does no affect the extremization process of the hypersurface. This factor accounts for the complexity associated to the internal degrees of freedom encoded in the compact part of the space, not included in the effective five dimensional treatment, that in this particular case turn out to be independent of the energy scale given by the radial coordinate. The explicit expression we obtain is
\begin{align}
\text{Vol}(\Sigma_{10}) = 2 \pi^3 V_x \int_{r_m}^{r_{\infty}} V \sqrt{W \left( \frac{1}{U}-U t'^2 \right) } dr,
\label{vol10}
\end{align}
where we have again factorized the volume $V_x$ coming from the gauge theory spatial directions exhibiting that this volume is equal to the one obtained in \eqref{5D_Vol1} times $\pi^3$. This is an interesting result, as it shows that in this particular case, the CV conjecture yields the same behavior using either the 5D truncation or the full ten dimensional background. This is explicitly seen by computing both $\Sigma_{up}$ and $\Sigma_{full}$.

We find $\Sigma_{up}$ by substituting in \eqref{vol10} the $t(r)$ obtained in the five dimensional case \eqref{t_5D}. Since this expression is proportional to \eqref{5D_Vol1}, we will find the same behavior as in five dimensions. On the other hand, we compute $\Sigma_{full}$ by looking for the $t(r)$ which extremizes \eqref{vol10}. However, we already know that the solution found in section \ref{CV_computation} is the one that takes \eqref{vol10} to its extremal value. We conclude that for the DK model, $\Sigma_{full} = \Sigma_{up}$ and, given that the volume of both is just the volume of $\Sigma$ times $\pi^3$, the results for the complexity in ten dimensions can be trivially read from the ones in five dimensions presented in SEC. \ref{results5D}. We will omit the corresponding plots as they provide no new information.

It is worth noticing that this is not a general behavior for the complexity, as exemplified by the AP model where $\Sigma_{full} \neq \Sigma_{Up}$. This is because in the line element \eqref{10DAP} the compact and non-compact directions mix in a non-trivial manner, preventing the dependence on $\theta$ in particular from being integrated out, and making an exclusively $r$ dependent embedding not general enough to reach a true extremal value for the volume of $\Sigma_{10}$. Thus, the parametrization cannot be the same as in the DK model. In its place, we choose $\xi^{a} = (r,x,y,z,\theta, \psi,\phi_1,\phi_2, \phi_3)$ and $x^{\mu}(\xi^{a}) = (t(r,\theta),r,x,y,z,\theta, \psi, \phi_{1},\phi_{2}, \phi_{3})$, noticing that now $t$ is a function of $r$ and $\theta$. With this selection we obtain 
\begin{align}
\mathcal{L}_{10}= \Lambda V(r) \sqrt{W(r)}\sqrt{\sqrt{\Delta }\frac{\left(1-t_r^2 U^2\right)-t_{\theta}^2 U X^{-1}}{U(r)}},
\label{lagrangian_10D_AP}
\end{align} 
where $t_r$ and $t_{\theta}$ respectively represent the derivatives of $t$ with respect to $r$ and $\theta$. We see that the expression for $\mathcal{L}_{10}$ in the AP model is relevantly different from the one in the DK model \eqref{5D_Vol1}. As anticipated, there is now an explicit dependence of the compact direction $\theta$ inside the integral, that in general cannot be integrated on its own as in \eqref{vol10}. In view of the above, we cannot expect the ten dimensional behavior to be the same as the five dimensional one, and computing both $\Sigma_{up}$ and $\Sigma_{full}$ will exhibit the details in the discrepancy. In order to compute $\Sigma_{up}$, we uplift the hypersurface obtained in the consistent truncation, substituting our five dimensional solution for $t(r)$, given by \eqref{t_5D}, resulting in the following volume
\begin{align}
\text{Vol}(\Sigma_{up})=2V_{x}\int_{r_{m}}^{r_{\infty}}F(X)\frac{V(r)^{2}W(r)}{\sqrt{E^{2}+U(r)V^{2}(r)W(r)}}dr,
\label{AP10_up}
\end{align}
where
\begin{equation}
F(X)=\frac{8\pi^{3}(4+X^{15/4}(5X^{3}-9))}{45 X^{1/2}(X^{3}-1)^{2}}.
\end{equation}

Determining $\Sigma_{full}$ requires the extremization of $\text{Vol}(\Sigma_{10})$, however, as can be seen in Appendix \ref{appendixEq}, the partial differential equation for the embedding that appears as part of the process is non-linear, second order, not-separable, and has alluded all our efforts, analytic, numeric, or hybrid, to solve it. Nonetheless, it seems like the physical conclusion that matters the most is that the current background is such that  $\Sigma_{full}$ differs from $\Sigma_{up}$. To prove this it suffice to postulate $t(t,\theta)$ as a function $t(r)$ of $r$ alone, which reduces the embedding equation to
\begin{align}
0 = & 12 \Delta  \text{EoM}_5+  \nonumber \\
 & \sqrt{6} U V W t' \varphi'  \left(2 \Delta -3 X \cos ^2(\theta )\right) \left(U^2 t'^2-1\right),
\label{AP10_full}
\end{align}
where $\text{EoM}_5$ can be read from \eqref{5D_EOM} as the quantity that must vanish to satisfy the equation of motion in the five dimensional case. We see now that the five dimensional solution given by $\text{EoM}_5=0$ would only solve the ten dimensional equation if either $t' = 0$ or $t' = 1/U$, which is respectively equivalent to taking $E = 0$ or $E \rightarrow \infty $ in equation \eqref{t_5D}. According to the CV conjecture, neither of these two solutions lead to hypersurfaces that can be used to compute the complexity, since the one described by $t' = 0$ does not connect the two boundaries smoothly, except for the very particular case $t=0\Rightarrow\tau=0$, while the one generated by $t' = 1/U$ is null and therefore fails the requirement to be space-like. This explicitly shows that in the AP model $\Sigma_{full}$ cannot be simply obtained by uplifting the five dimensional result, making it different from $\Sigma_{up}$, and leading us to conclude that the complexity computed with one of these two hypersurfaces will not coincide with the one that results from using the other, except for $\tau=0$.

Furthermore, and from a wider perspective, since $\Delta$ and $\cos^2(\theta)$ are linearly independent as functions of $\theta$, Eq. \eqref{AP10_full} shows that the only functions of $r$ alone that solve the complete embedding equation are those we already mention, and therefore any other must also depend on $\theta$. Finding this larger family of solutions is beyond the scope of this paper, and consequently in what follows we will limit our analysis to the volume of the $\Sigma_{up}$ hypersurfaces.

\subsection{AP model: results for $\Sigma_{up}$}

\begin{figure}[ht!]
 \centering
 \includegraphics[width=0.4\textwidth]{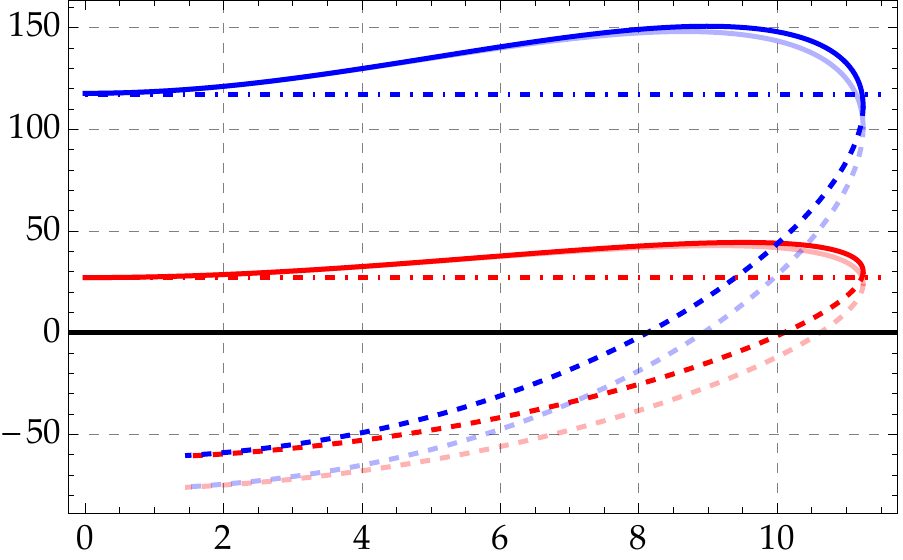}
\put(-230,110){\large $\frac{G_{N}\mathcal{C}_{E}}{V_{x}\pi^{3}T^{3}}$}
\put(0,-10){\large $\frac{B}{T^2}$}
\caption{\small Evolving complexity $G_{N}\mathcal{C}_{E}/V_{x}\pi^{3}T^{3}$ for the AP model in ten dimensions using $\Sigma_{up}$ and reported as a function of $B/T^2$ with $T \tau =0$ (red, bottom) and $T \tau =1$ (blue, top). The solid lines represent the stable branches, while the dashed ones are the unstable branches. The transparent lines show the results previously found using the five dimensional truncation. The horizontal dot-dashed lines denotes $\mathcal{C}_{E}$ for $B/T^{2}=0$ but the same $T\tau$ as the corresponding color.}
\label{CF_AP_10}
\end{figure}

The numerical calculations in this case show once again that the dimensionless ratio $\mathcal{C}/T^3$ depends only on the two dimensionless parameters $B/T^2$ and $T \tau$, indicating that our results are still insensitive to the conformal anomaly. To do any further comparison it is necessary to divide out the factor of $\pi^3$ by which the volume of the ten and five dimensional hypersurfaces differ at $B=0$, and that, as previously stated, is associated with the complexity of the internal degrees of freedom encoded in the volume of the compact dimensions. In FIG. \ref{CF_AP_10} we plot the evolving complexity computed using the volume of $\Sigma_{up}$ in the ten dimensional AP model once this scaling has been done. We see that despite the small, but existing, quantitative differences, the general behavior is very similar to the one we obtained in the five dimensional treatment, included in FIG. \ref{CF_AP_10} as transparent lines. Just like in the five dimensional case, the evolving complexity displayed in FIG. \ref{CF_AP_10} increases with $B/T^{2}$ until it peaks at a value of this dimensionless parameter below the critical one, decreasing from that point onward. This means that for certain intensities, it is easier to start from the vacuum and create a state with a very intense magnetic field than a state with a lower one.

\begin{figure}[ht!]
 \centering
 \includegraphics[width=0.4\textwidth]{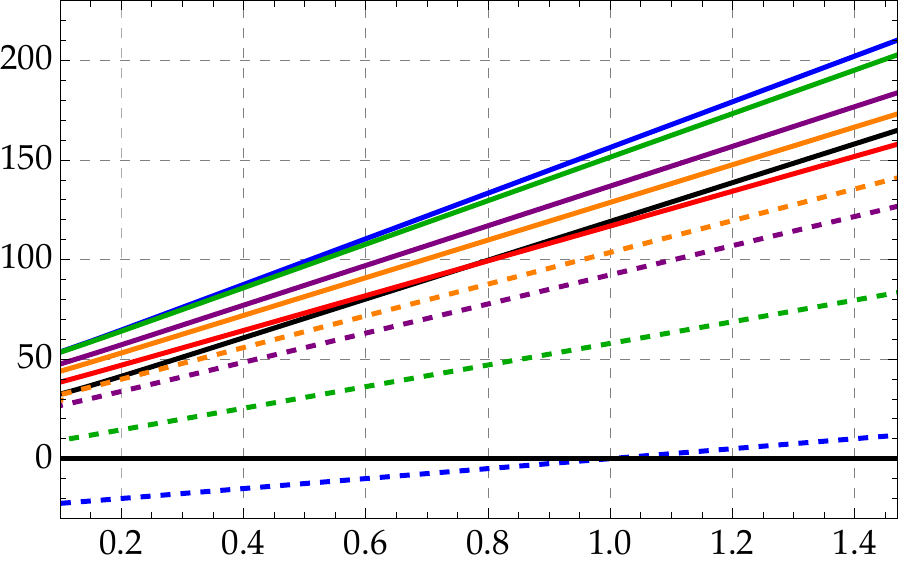}
\put(-230,100){\large $\frac{G_{N}\mathcal{C}_{E}}{V_{x}\pi^{3}T^{3}}$}
\put(0,-10){\large $T\tau$}
\caption{\small Evolving complexity $G_{N}\mathcal{C}_{E}/V_{x}\pi^{3}T^{3}$ for the AP model in ten dimensions using $\Sigma_{up}$ and reported as a function of $T\tau$ with $B/T^2 =  0$ as reference (black, fifth line down), and then (from top to bottom for solid lines and bottom to top for dashed ones), $B/T^2 \approx  8.08$ (blue), $B/T^2 \approx  10.37$ (green), $B/T^2 \approx  11.04$ (purple), $B/T^2 \approx  11.18$ (orange) and $B/T^2 \approx  11.24$ (red). The solid lines represent the stable branches, while the dashed ones are the unstable branches.}
\label{CF_AP_10_tau}
\end{figure}

In FIG. \ref{CF_AP_10_tau} we explore the behavior of the evolving complexity while keeping $B/T^{2}$ fixed. From this it can be seen that $C_{E}$ is a monotonically increasing function of $T\tau$ and that when $T \tau$ goes to infinity it does it at a constant rate. While this general behavior is the same for all the explored values of  $B/T^2$, it is important to note that for large enough $B/T^{2}$ the evolving complexity is smaller than the one obtained at vanishing magnetic field for every value of $T \tau$ that we checked, that is $\mathcal{C}_{E}(B/T^{2},T\tau)<\mathcal{C}_{E}(0,T\tau)$ for states that are part of the stable branch. To better present this effect it is again convenient to study the complexity of magnetization $\mathcal{C}_M$ that, as stated, isolates the magnetic contribution to the complexity.

\begin{figure}[ht!]
 \centering
 \includegraphics[width=0.4\textwidth]{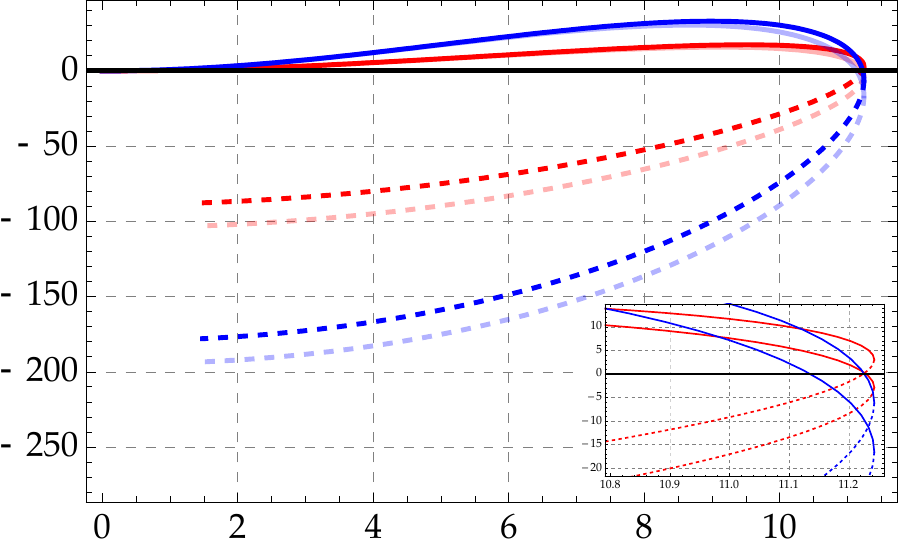}
\put(-220,120){\large $\frac{G_{N}\mathcal{C}_{M}}{V_{x}\pi^{3}T^{3}}$}
\put(0,-10){\large $\frac{B}{T^2}$}
\caption{\small Complexity of magnetization $G_{N}\mathcal{C}_{M}/V_{x}\pi^{3}T^{3}$ for the AP model in ten dimensions using $\Sigma_{up}$ and reported as a function of $B/T^2$ with $T \tau =0$ (red, closer to zero) and $T \tau =1$ (blue, farther from zero). The solid lines represent the stable branches, while the dashed ones are the unstable branches. The transparent lines show the results previously found using the five dimensional truncation. We present an inset of the large $B/T^{2}$ region in order to better visualize the magnetic simplification phenomenon.}
\label{CM_AP_10}
\end{figure}

As can be seen in FIG. \ref{CM_AP_10}, $\mathcal{C}_{M}$ at $T\tau=1$ becomes negative for values of the dimensionless ratio $B/T^2$ that are below the critical magnetic field intensity and are still part of the thermodynamically stable branch. This explicitly shows that forming a state in which the magnetic field has an intensity in this range of values is simpler than forming a state with the same physical parameters and no magnetic field. This is the magnetic simplification phenomenon that we encountered when studying the complexity in the consistent truncation of the theory: there exists a certain magnetic field intensity $B_{s}/T^2$ above which the complexity of magnetization becomes negative for a given $T\tau$. However, as can be appreciated in FIG. \ref{CM_AP_10}, $B_{s}$ for the 10-dimensional theory is larger than its five dimensional counterpart: a stronger magnetic field is necessary to reduce the complexity of the internal degrees of freedom appearing in the ten-dimensional scenario. This effect is such that, in contrast to what we found for the 5-dimensional truncated theory, for $T\tau=0$ there is no magnetic simplification phenomenon when working with the 10-dimensional theory.

\begin{figure}[ht!]
 \centering
 \includegraphics[width=0.4\textwidth]{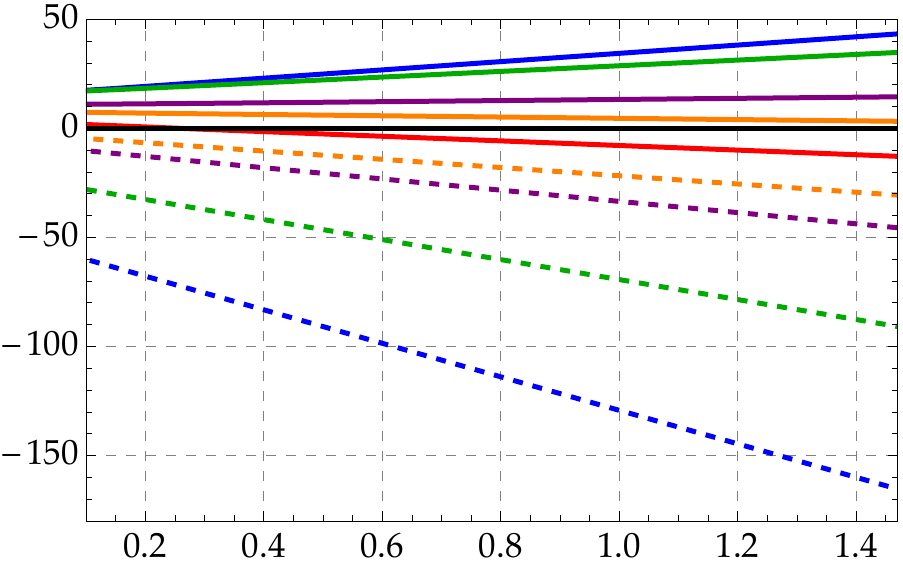}
\put(-220,110){\large $\frac{G_{N}\mathcal{C}_{M}}{V_{x}\pi^{3}T^{3}}$}
\put(0,-10){\large $T \tau$}
\caption{\small Complexity of magnetization $G_{N}\mathcal{C}_{M}/V_{x}\pi^{3}T^{3}$ for the AP model in ten dimensions using $\Sigma_{up}$ and reported as a function of $T \tau$, with $B/T^2 =  0$ (black, at zero) as a reference, and then (from top to bottom for solid lines and bottom to top for dashed ones), $B/T^2 \approx  8.08$ (blue), $B/T^2 \approx  10.37$ (green), $B/T^2 \approx  11.04$ (purple), $B/T^2 \approx  11.18$ (orange) and $B/T^2 \approx  11.24$ (red). The solid lines represent the stable branches, while the dashed ones are the unstable branches.}
\label{CM_AP_10_Bc}
\end{figure}

This behavior can be better appreciated in FIG. \ref{CM_AP_10_Bc}, where we plot $\mathcal{C}_{M}$ as a function of $T \tau$ for different values of $B/T^{2}$. We can see that, in contrast to what we found in the five-dimensional theory, the complexity of magnetization of the stable state with $B/T^2 \approx  11.18$ remains positive for $T\tau>0.6$, although it still becomes negative for late enough times. We also see that for small values of $B/T^2$, $\mathcal{C}_{M}$ for the states in the stable branch increases with $T \tau$. On the other hand, for high enough values of $B/T^2$ the behavior changes, and the complexity of magnetization decreases as $T \tau$ grows. It is important to clarify that the impression left by FIG. \ref{CM_AP_10_Bc} about how increasing $B/T^2$ will reduce the complexity for all $T \tau$ is due to the range of values for $B/T^2$, and was purposefully done to highlight the magnetic simplification phenomenon that is one of our more interesting results.

\section{Discussion}
\label{Discussion}

We computed the computational complexity using the CV conjecture for two different gravitational models dual to quantum field theories with a magnetic field, the D'Hoker-Kraus (DK) model and the Ávila-Patiño (AP) model, and for both contrasted the five dimensional effective version with the full ten dimensional theory.

As a first result, we verified that the evolving complexity and complexity of magnetization are both insensitive to the conformal anomaly present in these theories in both the full ten dimensional setup and its consistent truncation. We checked this by noticing that  $\mathcal{C}_{E}/T^3$ and $\mathcal{C}_{M}/T^3$ depend only on the dimensionless quantities $B/T^2$ and $T \tau$. This had already been proven to happen when the complexity is computed by means of the CA conjecture \cite{Avila:2021zhb}.

For the DK model, it was found that the evolving complexity of the state increases as the magnetic field intensity grows. This is not an unexpected result, as we can think of the evolving complexity as how difficult it is to prepare a certain state from a reference one at any given time. With this intuition, a state with a strong magnetic field should be more difficult to prepare as the desired field intensity reach higher values. However, for the AP model, the results were drastically different. The state becomes more complex as the magnetic field increases up to a value above which a phenomenon of 'magnetic simplification' occurs, meaning that the evolving complexity starts decreasing, reaching values even below the one at vanishing magnetic field.

To isolate the effect of the magnetic field on the complexity of a state and the aforementioned phenomenon, we introduced a new quantity that we term complexity of magnetization, $\mathcal{C}_{M}$, defined as the difference in the complexities of states that are identical to each other except for the presence of the magnetic field in one of them. One reason for this quantity to be useful is that it permits us to identify the states that have been magnetically simplified as those that satisfy $\mathcal{C}_{M} < 0$, which indeed happens for stable states with intensities of the magnetic field above a certain simplification value $B_{s}$.

The two systems we used to compute the complexity seem ideal to understand the origin of the phenomenon of magnetic simplification, since the only relevant difference is the presence of one extra scalar field in one of them. The observation that in the DK model, where no other field has a non-vanishing VEV, the complexity increases with the intensity of the magnetic field, confirms our understanding that a state with a more intense magnetic field would be more complex to prepare. In contrast, in the theory dual to the AP model where the magnetic simplification occurs, there is a single trace scalar operator of scaling dimension equal to 2, with a non vanishing VEV that, at fixed source, changes with the intensity of the magnetic filed. A coherent way to encompass the above is to ascribe the simplification to the scalar operator, in which case, what we present in FIGS. \ref{CM_AP_5} and \ref{CM_AP_10} is that for intensities below $B_{s}$, the complexity grows due to the magnetic field, but starting at $B_{s}$, this increment is smaller than the reduction due to the complexity of the scalar operator, with the accumulated effect of the latter eventually even surpassing the one of the former.

Concerning the comparison of the full ten dimensional theories and the effective five dimensional ones, we demonstrated that the results obtained from uplifting the extremal hypersurface found in 5D and those derived after following the extremization procedure en 10D, were the same in the DK model and different in the AP. Our results are not enough to support the use of $\Sigma_{up}$ or $\Sigma_{full}$ in the CV conjecture, but they certainly positions the dilemma as relevant.

\section*{Acknowledgments}

The work DA is partially supported by Mexico's National Council of Science and Technology (CONACyT) grant A1-S-22886 and DGAPA-UNAM grants IN107520 and IN116823 and additionally supported by a DGAPA-UNAM postdoctoral grant. CD is supported by CONACYT Ph. D grant. All the plots in this paper were generated using Wolfram Mathematica.
\appendix
\section{Interior solutions}
\label{AppInt}
In this appendix we show the integration procedure needed to obtain the interior solutions for the AP model. The treatment is analogous to the one for the DK model, which can be consulted in \cite{Avila:2021zhb}. The equations of motion for the metric, scalar and Maxwell fields come from the variation of the 5-dimensional truncated action \eqref{ReductionAction}. After substitution of the general anzats, the Maxwell equations are automatically satisfied and the Einstein and scalar equations can be manipulated as
\begin{equation}
\begin{split}
0= &2W^{2}(4B^{2}X^{-2}+V(U'V'+UV''))\\&-VW(2V(U'W'+UW'')+UV'W')+UV^{2}W'^{2}, \\
0= & 2W^{2}(V'^{2}-V(2V''+V\varphi'^2))-V^{2}(2W W''-W'^{2}), \\
0= & W(-8b^{2}X^{-2}+2V^{2}(3U''-8(X^{2}+2X^{-1}))\\&+ 6VU'V')+3V^{2}U'W', \\
0= & W(4\sqrt{2}B^{2}X^{-2}+V^{2}(2\sqrt{3}(U\varphi')'+8\sqrt{2}(X^{2}-X^{-1}))\\&+2\sqrt{3}UV\varphi'V')+\sqrt{3}UV^{2}\varphi'W', \\
0= & W(4B^{2}X^{-2}+2VU'V'+UV'^{2})\\&-WV^{2}(U\varphi'^{2}+8(X^{2}+2X^{-1}))\\&+VW'(VU'+2UV'),\\
\end{split}
\label{EOMAP}
\end{equation}

The first step is to numerically solve \eqref{EOMAP} is to expand them in powers of $r$ around $r_{h}$ using
\begin{eqnarray}
&& U=6r_{h}(r-r_{h})+\sum_{i=2}^{\infty}U_{i}(r-r_{h})^{i},
\cr
&& V=V_{0}+\sum_{i=1}^{\infty}V_{i}(r-r_{h})^{i},
\cr
&& W=3r_{h}^{2}+\sum_{i=1}^{\infty}W_{i}(r-r_{h})^{i},
\cr
&& \varphi=\varphi_{h}+\sum_{i=1}^{\infty}\varphi_{i}(r-r_{h})^{i}.
\label{HorizonAP}
\end{eqnarray}
This behavior near the horizon allows the family of solutions to easily interpolate between the D3-black brane for $B/V_{0}=0$ and $\varphi_{h}=0$ and the other members by changing the value of $B/V_{0}$ and $\varphi_{h}$. Additionally, this also ensures that the temperature of every member of the family is given by $T=3 r_{h}/2\pi$. 

Substitution of \eqref{HorizonAP} into \eqref{EOMAP} allows to solve for any of the undetermined coefficients in terms of $B/V_{0}$, and then use this to provide initial data for the numerical integration performed from $r=r_{h}+\epsilon$ to the boundary at $r=\infty$ for the exterior solutions, and from $r=r_{h}-\epsilon$ to the singularity at $r=r_{s}$ for the interior solutions, with $\epsilon\ll r_{h}$ in both cases. The boundary behavior of the solutions built with this procedure is
\begin{equation}
V\sim V_{bdry}r^{2}, \qquad W\sim W_{bdry}r^{2}, \qquad U\sim r^{2}.
\end{equation}
Nonetheless, we can exploit the symmetries of the equations of motion \eqref{EOMAP} to re-scale them as
\begin{equation}
V\rightarrow \frac{V}{V_{bdry}}, \qquad W\rightarrow \frac{W}{W_{bdry}}, \qquad B\rightarrow \frac{B}{V_{bdry}},
\end{equation}
which in turn gives the desired $AdS_{5}$ behavior at the boundary. Note that this re-scaling needs to be done consistently for both the exterior and interior solutions and that it is necessary to simultaneously scale the value of $B$ to preserve the solution. It is also important to mention that the position of the singularity is not fixed at $r_{s}=-r_{h}/2$ for every member of the family of solutions, but only for $B/T^{2}=0$. Instead, the location of the singularity in the $r$-coordinate turns out to be a function of the magnetic field intensity. By $r_{s}$ we mean the radius at which the curvature scalar $R_{\mu\nu\alpha\beta}R^{\mu\nu\alpha\beta}$ diverges. This behavior is shared for both the AP and DK models.

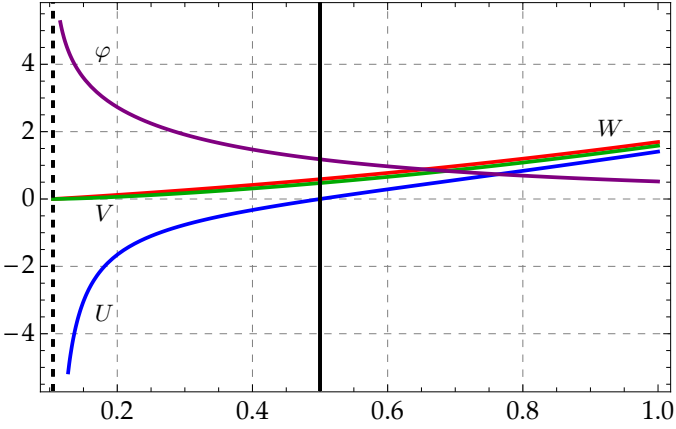
\begin{figure}[ht!]
 \centering
 \includegraphics[width=0.5\textwidth]{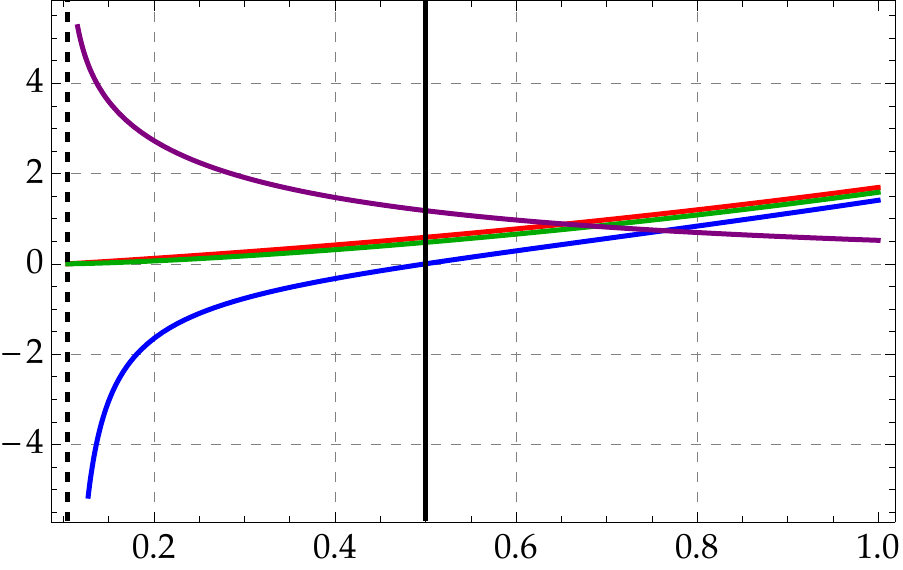}
\put(-220,40){$U$}
\put(-220,140){$\varphi$}
\put(-30,110){$W$}
\put(-220,78){$V$}
\caption{\small Metric functions for the AP model as functions of $r$ for $B/T^{2}=11.24$. The horizon, located at $r_{h}=1/2$, is denoted as a black vertical line while the singularity, located at $r_{s}=0.1$ for this magnetic field intensity, is denoted as a dashed black vertical line.}
\label{APBackgroundPlot}
\end{figure}

Now, the family of solutions found with the procedure just described depends on the three independent parameters $r_{h}$, $B/V_{0}$ and $\varphi_{h}$. This coincides with the number of free parameters from the perspective of the dual gauge theory: the temperature $T$ and the magnetic field intensity $B$ on the plasma, and the source of the scalar operator $\mathcal{O}_{\varphi}$ dual to $\varphi$. The latter is dual to the coefficient $\psi_{0}$ that appears in the boundary expansion of the scalar field
\begin{equation}
\varphi\rightarrow\frac{1}{r^{2}}\left(\varphi_{0}+\psi_{0}\log{r}\right).
\end{equation}
Given that from the perspective of the dual gauge theory it makes sense to work at a fixed $\psi_{0}$, in practice we solve \eqref{EOMAP} for different values of $r_{h}$, $B/V_{0}$ and $\varphi_{h}$ and then use that to numerically determine the value of $\varphi_{h}$ that fixes $\psi_{0}$ for any given $B$ and $T$. The family of solutions studied in the main text correspond to the one with the source term turned off $\psi_{0}=0$. We show the metric functions for the critical magnetic field $B/T^{2}=11.24$ in the interior and exterior regions in Fig. (\ref{APBackgroundPlot}).

\section{Penrose diagram}
\label{AppPenrose}
In this appendix we show how to construct the Penrose Diagram for the two sided black hole geometries studied in the main text. Starting from the general anzats for the line element \eqref{5DBulk}, we first change to the tortoise coordinate $r_{\star}$ given by the solution to the equation
\begin{equation}
\frac{dr_{\star}}{dr}=\frac{1}{U(r)},
\label{Tortoise}
\end{equation}
that satisfies the boundary condition $r_{\star}(\infty)=0$. Note that near the horizon we have
\begin{equation}
r_{\star}\simeq\frac{\log(r-r_{h})}{4\pi T},
\end{equation}
because of the behavior of the metric function $U(r)$ given in \eqref{HorizonAP}. Next we transform to the Kruskal-Szekeres coordinates, given by
\begin{eqnarray}
&& \mathcal{U}=+e^{-2\pi T(t-r_{\star})}, \quad \mathcal{V}=-e^{2\pi T(t+r_{\star})} \quad \text{Left exterior}
\cr
&& \mathcal{U}=-e^{-2\pi T(t-r_{\star})}, \quad \mathcal{V}=+e^{2\pi T(t+r_{\star})} \quad \text{Right exterior}
\cr
&& \mathcal{U}=+e^{-2\pi T(t-r_{\star})}, \quad \mathcal{V}=+e^{2\pi T(t+r_{\star})} \quad \text{Future interior}
\cr
&& \mathcal{U}=-e^{-2\pi T(t-r_{\star})}, \quad \mathcal{V}=-e^{2\pi T(t+r_{\star})} \quad \text{Past interior} \cr
&&
\end{eqnarray}
and finally we change to the compact coordinates
\begin{equation}
X=\frac{\arctan \mathcal{V}-\arctan \mathcal{U}}{2}, \qquad Y=\frac{\arctan \mathcal{V}+\arctan \mathcal{U}}{2},
\end{equation}
which are globally spacelike and timelike respectively. This are the coordinates in which we plot the Penrose Diagram in FIG. \eqref{Penrose} presented in the main text.

\section{Boundary expansions}
\label{AppBdry}
The near boundary behavior of the geometries that are part of the DK or AP models can be obtained by solving the equations of motion coming from \eqref{ReductionAction} as a power series in $r$ around infinity. The only restrictions we impose is that the metric asymptotes exactly the metric of $AdS_{5}$ and, in the case of the AP model, that the non-normalizable mode of the scalar field is turned off. The result for the DK model reads
\begin{align}
U(r)=& r^{2}+U_{1}r+\frac{U_{1}^{2}}{4}+\frac{1}{r^{2}}\left(U_{4}-\frac{2}{3}b^{2}\log{r}\right) \nonumber \\&+\mathcal{O}\left(\frac{1}{r^{4}}\right),\\
V(r)=& r^{2}+U_{1}r+\frac{U_{1}^{2}}{4} \nonumber \\&+\frac{1}{r^{2}}\left(-\frac{1}{2}W_{4}+\frac{1}{3}b^{2}\log{r}\right)+\mathcal{O}\left(\frac{1}{r^{4}}\right),\\
W(r)=& r^{2}+U_{1}r+\frac{U_{1}^{2}}{4}+\frac{1}{r^{2}}\left(W_{4}-\frac{2}{3}b^{2}\log{r}\right) \nonumber \\&+\mathcal{O}\left(\frac{1}{r^{4}}\right),
\label{DK_r_expansions}
\end{align}
while for the AP model we have that
\begin{equation}
\begin{aligned}
U(r)=& r^{2}+U_{1}r+\frac{U_{1}^{2}}{4}+\frac{1}{r^{2}}\left(U_{4}-\frac{2}{3}b^{2}\log{r}\right)\\&+\mathcal{O}\left(\frac{1}{r^{4}}\right),\\
V(r)=& r^{2}+U_{1}r+\frac{U_{1}^{2}}{4}\\&+\frac{1}{r^{2}}\left(-\frac{1}{2}W_{4}-\frac{1}{6}\varphi_{0}^{2}+\frac{1}{3}b^{2}\log{r}\right)+\mathcal{O}\left(\frac{1}{r^{4}}\right),\\
W(r)=& r^{2}+U_{1}r+\frac{U_{1}^{2}}{4}+\frac{1}{r^{2}}\left(W_{4}-\frac{2}{3}b^{2}\log{r}\right)\\&+\mathcal{O}\left(\frac{1}{r^{4}}\right),\\
\varphi(r)=& \frac{\varphi_{0}}{r^{2}}-\frac{U_{1}\varphi_{0}}{r^{3}}+\frac{1}{12r^{4}}\left(-2\sqrt{6}b^{2}\right.\\&\left.+\varphi_{0}(9U_{1}^{2}-\sqrt{6}\varphi_{0})\right)+\mathcal{O}\left(\frac{1}{r^{5}}\right),
\end{aligned}
\label{AP_r_expansions}
\end{equation}
where $U_{1}$, $U_{4}$, $W_{4}$ and $\varphi_{0}$ are coefficients that are not determined by the equations of motion, but can be read as functions of the magnetic field intensity $b$ and the temperature $T$ once a particular numerical solution is known. Physically, $\varphi_{0}$ is dual to the VEV of the scalar operator $\langle\mathcal{O}_{\varphi}\rangle$ while $U_{4}$ and $W_{4}$ are both related to the stress-energy tensor \cite{Avila:2018hsi,Avila:2021zhb}.

\section{Rate of change of the complexity} 
\label{dCdtau}
In this appendix we present the computation of the rate of change of the complexity $d\mathcal{C}/d\tau$ for the five-dimensional models studied in the main text. We begin by noting that by means of \eqref{t_5D} we can relate $\tau$ and $r_{m}(\tau)$ implicitly by
\begin{equation}
\begin{aligned}
\tau&=\int^{r_{\infty}}_{r_{m}(\tau)}t'(r)dr\\
&=\int^{r_{\infty}}_{r_{m}(\tau)}\frac{E(\tau)}{\sqrt{E(\tau)^{2}+U(r)V^{2}(r)W(r)}}dr
\end{aligned}
\label{tauApp}
\end{equation}
as by definition $t(r_{\infty})=\tau$ and $t(r_{m})=0$. 

Using the previous expression we can rewrite \eqref{5D_Vol} as
\begin{equation}
\begin{aligned}
\frac{\text{Vol}(\Sigma)}{2V_{x}}=&\int_{r_{m}(\tau)}^{r_{\infty}}\frac{\sqrt{E(\tau)^{2}+U(r)V^{2}(r)W(r)}}{U(r)}-E(\tau)\tau,
\end{aligned}
\end{equation}
which is suitable to compute the derivative of the volume with respect to $\tau$. Indeed, a direct computation yields
\begin{equation}
\begin{aligned}
&\frac{1}{2V_{x}}\frac{d\text{Vol}(\Sigma)}{d\tau}=-\frac{\sqrt{E(\tau)^{2}+U(r_{m})V^{2}(r_{m})W(r_{m})}}{U(r_{m})}\frac{dr_{m}}{d\tau}\\&
+\frac{dE}{d\tau}\left(\int_{r_{m}(\tau)}^{r_{\infty}}\frac{E(\tau)}{\sqrt{E(\tau)^{2}+U(r)V^{2}(r)W(r)}}dr-\tau\right)-E.
\end{aligned}
\end{equation}
The first term vanishes by the definition of $E$ given in \eqref{ENS}, while the second does by virtue of \eqref{tauApp}. Hence we are left with
\begin{equation}
\frac{d\mathcal{C}_{V}}{d\tau}=-\frac{2V_{x}}{G_{N}}E(\tau)=\frac{2V_{x}}{G_{N}}U(r_{m})W(r_{m})V(r_{m})^{2}.
\end{equation}

This expression holds for any value of the boundary time $\tau$. However, in the limit $\tau\rightarrow\infty$ we have that $r_{m}\rightarrow r_{min}$, with $r_{min}$ defined in \eqref{rmin}. Thus the late time behavior of the rate of change of the complexity is given by
\begin{equation}
\lim_{\tau\rightarrow\infty}\frac{d\mathcal{C}_{V}}{d\tau}=\frac{2V_{x}}{G_{N}}U(r_{min})W(r_{min})V(r_{min})^{2}.
\label{latetimes}
\end{equation}
For $B=0$, which corresponds to the Black D3-brane solution, equation \eqref{rmin} can be solved analytically. For this specific case we have
\begin{equation}
\lim_{\tau\rightarrow\infty}\frac{d\mathcal{C}_{V}}{d\tau}=\frac{V_{x}}{G_{N}}\pi^{2}T^{4},
\end{equation}
which corresponds to the expected constant behavior consistent with Lloyd's bound \cite{Susskind:2014rva,Stanford:2014jda,Carmi:2017jqz} as the energy density of the state at temperature $T$ is proportional to $T^{4}$. 

Given that there is no analytical way to compute $r_{min}$ for any of the solutions with $B\neq 0$ in either the DK and AP models, equation \eqref{latetimes} needs to be evaluated numerically. In all the cases that we explored, said evaluation revealed that $\mathcal{C}_{V}$ grows at a constant rate as $\tau\rightarrow\infty$, which is consistent with the results presented in the main text. Given that the same conclusion was obtained by two different methods, this gives confirmation on the validity of our numerical procedures. The question of whether Lloyd's bound is satisfied for states at $B\neq 0$ in the DK or AP models is a complicated one, as because of the conformal anomaly present in both the specification of the energy density requires fixing a renormalization scheme (see \cite{Avila:2018hsi,Ballon-Bayona:2022uyy}). We have previously showed that, when working using the CA holographic prescription, it is possible to use the saturation of Lloyd's bound at late times to fix said renormalization scheme for the DK and Mateos-Trancanelli models. We expect that the same conclusion also apply for the CV prescription.

\section{$\Sigma_{full}$ equations} 
\label{appendixEq}
It was shown in section \ref{Complexity_10D} that in the AP model, $\Sigma_{up}$ and $\Sigma_{full}$ are different hypersurfaces except for certain particular cases. We claimed back then that in general $\Sigma_{full}$ is given by an embedding function $t(r,\theta)$ that necessarily depends on both coordinates $r$ and $\theta$, and we will see now that assuming a sole dependence in $r$ is inconsistent with the embedding equation. To prove this, we need to extremize the volume of an hypersurface described by $t(r, \theta)$, that connects the boundaries dual to both theories in the double thermo-field setup for the full ten dimentional theory. Such volume is given by
\begin{align}
\text{Vol} (\Sigma_{full}) = 2 \int \mathcal{L}_{10} d^9x
\end{align}
where $\mathcal{L}_{10}$ is defined by equation \eqref{lagrangian_10D_AP}, and its variation with respect to the embedding function results in the partial non-linear differential equation
\begin{widetext}
\begin{align}
0 = & 6 \Delta  \sin (2 \theta ) U e^{\sqrt{\frac{3}{2}} \varphi} t_{r} \left(2 W V'+V W'\right) \left(U^2 e^{\frac{\varphi}{\sqrt{6}}} t_{r}^2+U t_{\theta}^2-e^{\frac{\varphi}{\sqrt{6}}}\right) \nonumber \\
 & +3 U V W t_{\theta} \left(6 \sin ^2(\theta ) (3 \cos (2 \theta )-1)+2 (9 \cos (2 \theta )-5) \cos ^2(\theta ) e^{\sqrt{\frac{3}{2}} \varphi}\right) \left(t_{\theta}^2+U e^{\frac{\varphi}{\sqrt{6}}} t_{r}^2\right) \nonumber \\
& + 3 \sin (2 \theta ) e^{\sqrt{\frac{3}{2}} \varphi} V W t_{\theta} \left(7 \sin (2 \theta ) e^{\frac{\varphi}{\sqrt{6}}}-8 \Delta  U^2 t_{r} t_{r \theta}\right)+6 \sin ^2(\theta ) (1-3 \cos (2 \theta ))-8 \cos ^4(\theta ) e^{\sqrt{\frac{3}{2}} \varphi} \nonumber \\
 & +3 \sin (\theta ) \cos (\theta ) U e^{\sqrt{\frac{3}{2}} \varphi} V W t_{\theta}^2 \left(8 \Delta  U' t_{r}+U \left(8 \Delta  t_{rr}+\sqrt{6} \cos ^2(\theta ) e^{\frac{\varphi}{\sqrt{6}}} \varphi' t_{r}\right)\right) \nonumber \\
 & +12 \Delta  e^{\sqrt{\frac{2}{3}} \varphi} \left(U' t_{r} \left(U^2 t_{r}^2-3\right)+\sin (2 \theta ) e^{\frac{\varphi}{\sqrt{6}}} V W t_{\theta \theta} \left(U^2 t_{r}^2-1\right)\right) \nonumber \\
& +\sin (\theta ) \cos (\theta ) U e^{\sqrt{\frac{2}{3}} \varphi} V W \left(\sqrt{6} \varphi' t_{r} \left(U^2 t_{r}^2-1\right) \left(\cos ^2(\theta ) e^{\sqrt{\frac{3}{2}} \varphi}-2 \sin ^2(\theta )\right)-24 \Delta  e^{\sqrt{\frac{2}{3}} \varphi} t_{rr}\right),
\end{align}
\end{widetext}
where $U,V,W \varphi$ are functions of the coordinate $r$ alone, the primes represent the derivative with respect to $r$, the wrapping factor $\Delta$ is the function of both $r$ and $\theta$ defined by \eqref{wrapping}, and 
\begin{align}
& t_{r} = \frac{\partial t}{\partial r}; &  &t_{\theta} = \frac{\partial t}{\partial \theta}; \nonumber \\
& t_{rr} = \frac{\partial^2 t}{\partial r^2} & & t_{\theta \theta} = \frac{\partial^2 t}{\partial \theta^2}; \nonumber \\
& t_{r \theta} = \frac{\partial^2 t}{\partial r \partial \theta}.
\end{align}

We see by direct substitution that setting to zero all derivatives of $t$ with respect to $\theta$ in the above leads to an inconsistent equation except for the two cases listed in section \ref{Complexity_10D}.

\newpage
\bibliography{Complexity.bib}

\begin{thebibliography}{10}

\bibitem{Maldacena:1997re}
Juan~Martin Maldacena.
\newblock {The Large N limit of superconformal field theories and
  supergravity}.
\newblock {\em Int. J. Theor. Phys.}, 38:1113--1133, 1999.
\newblock [Adv. Theor. Math. Phys.2,231(1998)].

\bibitem{Ryu:2006bv}
Shinsei Ryu and Tadashi Takayanagi.
\newblock {Holographic derivation of entanglement entropy from AdS/CFT}.
\newblock {\em Phys. Rev. Lett.}, 96:181602, 2006.

\bibitem{Hubeny:2007xt}
Veronika~E. Hubeny, Mukund Rangamani, and Tadashi Takayanagi.
\newblock {A Covariant holographic entanglement entropy proposal}.
\newblock {\em JHEP}, 07:062, 2007.

\bibitem{Lewkowycz:2013nqa}
Aitor Lewkowycz and Juan Maldacena.
\newblock {Generalized gravitational entropy}.
\newblock {\em JHEP}, 08:090, 2013.

\bibitem{Dong:2016hjy}
Xi~Dong, Aitor Lewkowycz, and Mukund Rangamani.
\newblock {Deriving covariant holographic entanglement}.
\newblock {\em JHEP}, 11:028, 2016.

\bibitem{Takayanagi:2017knl}
Tadashi Takayanagi and Koji Umemoto.
\newblock {Entanglement of purification through holographic duality}.
\newblock {\em Nature Phys.}, 14(6):573--577, 2018.

\bibitem{Susskind:2014rva}
Leonard Susskind.
\newblock {Computational Complexity and Black Hole Horizons}.
\newblock {\em Fortsch. Phys.}, 64:24--43, 2016.
\newblock [Addendum: Fortsch.Phys. 64, 44--48 (2016)].

\bibitem{Stanford:2014jda}
Douglas Stanford and Leonard Susskind.
\newblock {Complexity and Shock Wave Geometries}.
\newblock {\em Phys. Rev. D}, 90(12):126007, 2014.

\bibitem{Brown:2015bva}
Adam~R. Brown, Daniel~A. Roberts, Leonard Susskind, Brian Swingle, and Ying
  Zhao.
\newblock {Holographic Complexity Equals Bulk Action?}
\newblock {\em Phys. Rev. Lett.}, 116(19):191301, 2016.

\bibitem{Brown:2015lvg}
Adam~R. Brown, Daniel~A. Roberts, Leonard Susskind, Brian Swingle, and Ying
  Zhao.
\newblock {Complexity, action, and black holes}.
\newblock {\em Phys. Rev. D}, 93(8):086006, 2016.

\bibitem{Carmi:2016wjl}
Dean Carmi, Robert~C. Myers, and Pratik Rath.
\newblock {Comments on Holographic Complexity}.
\newblock {\em JHEP}, 03:118, 2017.

\bibitem{Belin:2021bga}
Alexandre Belin, Robert~C. Myers, Shan-Ming Ruan, G\'abor S\'arosi, and
  Antony~J. Speranza.
\newblock {Does Complexity Equal Anything?}
\newblock {\em Phys. Rev. Lett.}, 128(8):081602, 2022.

\bibitem{Belin:2022xmt}
Alexandre Belin, Robert~C. Myers, Shan-Ming Ruan, G\'abor S\'arosi, and
  Antony~J. Speranza.
\newblock {Complexity Equals Anything II}.
\newblock 10 2022.

\bibitem{Carmi:2017jqz}
Dean Carmi, Shira Chapman, Hugo Marrochio, Robert~C. Myers, and Sotaro
  Sugishita.
\newblock {On the Time Dependence of Holographic Complexity}.
\newblock {\em JHEP}, 11:188, 2017.

\bibitem{Swingle:2017zcd}
Brian Swingle and Yixu Wang.
\newblock {Holographic Complexity of Einstein-Maxwell-Dilaton Gravity}.
\newblock {\em JHEP}, 09:106, 2018.

\bibitem{Mahapatra:2018gig}
Subhash Mahapatra and Pratim Roy.
\newblock {On the time dependence of holographic complexity in a dynamical
  Einstein-dilaton model}.
\newblock {\em JHEP}, 11:138, 2018.

\bibitem{Alishahiha:2018tep}
Mohsen Alishahiha, Amin Faraji~Astaneh, M.~Reza Mohammadi~Mozaffar, and Ali
  Mollabashi.
\newblock {Complexity Growth with Lifshitz Scaling and Hyperscaling Violation}.
\newblock {\em JHEP}, 07:042, 2018.

\bibitem{HosseiniMansoori:2018gdu}
Seyed~Ali Hosseini~Mansoori, Viktor Jahnke, Mohammad~M. Qaemmaqami, and
  Yaithd~D. Olivas.
\newblock {Holographic complexity of anisotropic black branes}.
\newblock {\em Phys. Rev. D}, 100(4):046014, 2019.

\bibitem{Auzzi:2022bfd}
Roberto Auzzi, Stefano Bolognesi, Eliezer Rabinovici, Fidel~I.
  Schaposnik~Massolo, and Gianni Tallarita.
\newblock {On the time dependence of holographic complexity for charged AdS
  black holes with scalar hair}.
\newblock {\em JHEP}, 08:235, 2022.

\bibitem{LloydBound}
S.~Lloyd.
\newblock {Ultimate physical limits to computation}.
\newblock {\em Nature}, 406:1047–1054, 2000.

\bibitem{Reynolds:2016rvl}
Alan Reynolds and Simon~F. Ross.
\newblock {Divergences in Holographic Complexity}.
\newblock {\em Class. Quant. Grav.}, 34(10):105004, 2017.

\bibitem{Kim:2017lrw}
Run-Qiu Yang, Chao Niu, and Keun-Young Kim.
\newblock {Surface Counterterms and Regularized Holographic Complexity}.
\newblock {\em JHEP}, 09:042, 2017.

\bibitem{Emparan:2021hyr}
Roberto Emparan, Antonia~Micol Frassino, Martin Sasieta, and Marija
  Toma\v{s}evi\'c.
\newblock {Holographic complexity of quantum black holes}.
\newblock {\em JHEP}, 02:204, 2022.

\bibitem{Couch:2017yil}
Josiah Couch, Stefan Eccles, Willy Fischler, and Ming-Lei Xiao.
\newblock {Holographic complexity and noncommutative gauge theory}.
\newblock {\em JHEP}, 03:108, 2018.

\bibitem{Avila:2020ved}
Daniel \'Avila and Leonardo Pati\~no.
\newblock {Melting holographic mesons by cooling a magnetized quark gluon
  plasma}.
\newblock {\em JHEP}, 06:010, 2020.

\bibitem{Chapman:2016hwi}
Shira Chapman, Hugo Marrochio, and Robert~C. Myers.
\newblock {Complexity of Formation in Holography}.
\newblock {\em JHEP}, 01:062, 2017.

\bibitem{Maldacena:2001kr}
Juan~Martin Maldacena.
\newblock {Eternal black holes in anti-de Sitter}.
\newblock {\em JHEP}, 04:021, 2003.

\bibitem{Fu:2018kcp}
Zicao Fu, Alexander Maloney, Donald Marolf, Henry Maxfield, and Zhencheng Wang.
\newblock {Holographic complexity is nonlocal}.
\newblock {\em JHEP}, 02:072, 2018.

\bibitem{Chapman:2018bqj}
Shira Chapman, Dongsheng Ge, and Giuseppe Policastro.
\newblock {Holographic Complexity for Defects Distinguishes Action from
  Volume}.
\newblock {\em JHEP}, 05:049, 2019.

\bibitem{Flory:2018akz}
Mario Flory and Nina Miekley.
\newblock {Complexity change under conformal transformations in
  AdS$_{3}$/CFT$_{2}$}.
\newblock {\em JHEP}, 05:003, 2019.

\bibitem{Bernamonti:2020bcf}
Alice Bernamonti, Federico Galli, Juan Hernandez, Robert~C. Myers, Shan-Ming
  Ruan, and Joan Sim\'on.
\newblock {Aspects of The First Law of Complexity}.
\newblock {\em J. Phys. A}, 53:29, 2020.

\bibitem{Engelhardt:2021mju}
Netta Engelhardt and \r{A}smund Folkestad.
\newblock {General bounds on holographic complexity}.
\newblock {\em JHEP}, 01:040, 2022.

\bibitem{Engelhardt:2021kyp}
Netta Engelhardt and \r{A}smund Folkestad.
\newblock {Negative complexity of formation: the compact dimensions strike
  back}.
\newblock {\em JHEP}, 07:031, 2022.

\bibitem{Cvetic:1999xp}
Mirjam Cvetic, M.~J. Duff, P.~Hoxha, James~T. Liu, Hong Lu, J.~X. Lu,
  R.~Martinez-Acosta, C.~N. Pope, H.~Sati, and Tuan~A. Tran.
\newblock {Embedding AdS black holes in ten-dimensions and eleven-dimensions}.
\newblock {\em Nucl. Phys.}, B558:96--126, 1999.

\bibitem{DHoker:2009mmn}
Eric D'Hoker and Per Kraus.
\newblock {Magnetic Brane Solutions in AdS}.
\newblock {\em JHEP}, 10:088, 2009.

\bibitem{Avila:2018hsi}
Daniel \'Avila and Leonardo Pati\~no.
\newblock {Instability of a magnetized QGP sourced by a scalar operator}.
\newblock {\em JHEP}, 04:086, 2019.

\bibitem{LachiezeRey:2005hs}
Marc Lachieze-Rey and S.~Caillerie.
\newblock {Laplacian eigenmodes for spherical spaces}.
\newblock {\em Class. Quant. Grav.}, 22:695--708, 2005.

\bibitem{Achour:2015zpa}
J.~Ben~Achour, E.~Huguet, J.~Queva, and J.~Renaud.
\newblock {Explicit vector spherical harmonics on the 3-sphere}.
\newblock {\em J. Math. Phys.}, 57(2):023504, 2016.

\bibitem{Avila:2021zhb}
Daniel \'Avila, C\'esar D\'\i{}az, Yaithd~D. Olivas, and Leonardo Pati\~no.
\newblock {Insensitivity of the complexity rate of change to the conformal
  anomaly and Lloyd\textquoteright{}s bound as a possible renormalization
  condition}.
\newblock {\em Phys. Rev. D}, 104(6):066011, 2021.

\bibitem{Arean:2016het}
Daniel Are\'an, Leopoldo~A. Pando~Zayas, Leonardo Pati\~no, and Mario
  Villasante.
\newblock {Velocity Statistics in Holographic Fluids: Magnetized Quark-Gluon
  Plasma and Superfluid Flow}.
\newblock {\em JHEP}, 10:158, 2016.

\bibitem{Avila:2018sqf}
Daniel Avila, Viktor Jahnke, and Leonardo Patiño.
\newblock {Chaos, Diffusivity, and Spreading of Entanglement in Magnetic
  Branes, and the Strengthening of the Internal Interaction}.
\newblock {\em JHEP}, 09:131, 2018.

\bibitem{Breitenlohner:1982jf}
Peter Breitenlohner and Daniel~Z. Freedman.
\newblock {Stability in Gauged Extended Supergravity}.
\newblock {\em Annals Phys.}, 144:249, 1982.

\bibitem{Bianchi:2001kw}
Massimo Bianchi, Daniel~Z. Freedman, and Kostas Skenderis.
\newblock {Holographic renormalization}.
\newblock {\em Nucl. Phys. B}, 631:159--194, 2002.

\bibitem{Avila:2019pua}
Daniel \'Avila and Leonardo Pati\~no.
\newblock {Melting holographic mesons by applying a magnetic field}.
\newblock {\em Phys. Lett. B}, 795:689--693, 2019.

\bibitem{Ballon-Bayona:2022uyy}
Alfonso Ballon-Bayona, Jonathan~P. Shock, and Dimitrios Zoakos.
\newblock {Magnetising the $ \mathcal{N} $ = 4 Super Yang-Mills plasma}.
\newblock {\em JHEP}, 06:154, 2022.

\end{thebibliography}
\bibliographystyle{unsrt}
\end{document}